\documentclass{aa}  

\usepackage{graphicx}
\usepackage{txfonts}
\usepackage{lipsum}
\usepackage{subcaption}
\usepackage{xcolor}

\usepackage{lscape}             
\usepackage{placeins}           
\usepackage{hyperref}      

\usepackage{cuted}
\usepackage{capt-of}

\emergencystretch=\maxdimen     
\begin{document}

   \title{The LOFAR Decametre Sky Survey (LoDeSS)}

   \subtitle{I. Survey description and first data release}

   \author{C.M. Cordun\inst{1,2}\fnmsep\thanks{Corresponding author: cordun@astron.nl}
        \and C. Groeneveld\inst{3}
        \and H.W. Edler \inst{1} 
        \and H.K. Vedantham \inst{1,2}
        \and F. de Gasperin \inst{3,4}
        \and R.J. van Weeren \inst{5}
        \and J. R. Callingham \inst{1,6}
        \and E. Osinga \inst{5}
        \and F. F. S. van der Tak \inst{2,7}
        \and M. J. Hardcastle \inst{8}
        \and M. Brüggen \inst{4}
        \and A. Botteon \inst{3}
        \and H. Röttgering \inst{5}
        \and T.W. Shimwell \inst{1,5}
        \and M. Iacobelli \inst{1}
        }

   \institute{ASTRON, Netherlands Institute for Radio Astronomy, Oude Hoogeveensedijk 4, Dwingeloo 7991PD, The Netherlands
   \and Kapteyn Astronomical Institute, University of Groningen, PO Box 800, 9700 AV Groningen, The Netherlands
   \and INAF -- Istituto di Radioastronomia, via P. Gobetti 101, 40129 Bologna, Italy
   \and Hamburger Sternwarte, Universität Hamburg, Gojenbergsweg 112, 21029 Hamburg, Germany
   \and Leiden Observatory, Leiden University, Niels Bohrweg 2, 2333 CA Leiden, the Netherlands
   \and Anton Pannekoek Institute for Astronomy, University of Amsterdam, Science Park 904, 1098 XH Amsterdam, The Netherlands
   \and SRON Netherlands Institute for Space Research, Landleven 12, 9747AD Groningen, The Netherlands
   \and Centre for Astrophysics Research, Department of Physics, Astronomy and Mathematics, University of Hertfordshire, College Lane, Hatfield AL10 9AB, UK}
   \date{Received August 4, 2026 / Accepted September 23, 2026}

 \abstract{
The decametre radio sky ($\sim 10$--$30$ MHz) remains largely unexplored due to strong ionospheric distortions and severe radio-frequency interference. Despite its scientific potential, including access to steep-spectrum synchrotron emission from aged electron populations and coherent radio emission from planetary and stellar magnetospheres, sub-arcminute imaging at these frequencies has only recently become feasible with instruments such as the Low Frequency Array (LOFAR). In this work, we present the first data release (DR1) of the LOFAR decametre Sky Survey (LoDeSS), which maps the northern sky above a declination of $20^\circ$ between 15 and 30 MHz. LoDeSS reaches an angular resolution of $\sim45\arcsec$ and typical noise levels of $\sim11.5$ mJy beam$^{-1}$ at a central frequency of 23~MHz. Observations were obtained with the LOFAR Low Band Antenna (LBA) system using a multi-beam strategy, with typical integration times of $\sim5$ h per pointing. They were carried out at night to reduce the impact of ionospherically reflected radio-frequency interference. Data quality varies across fields as a result of the ionospheric conditions and the fraction of data flagged due to radio-frequency interference. This first release covers $\sim15\%$ of the full survey area and consists of fields originally selected around exoplanet candidates for radio auroral emission. The products released to the community include primary-beam-corrected images in Stokes I and V, and a source catalogue for fields with reliable flux density scales. This release demonstrates that LOFAR enables sensitive, wide-field imaging in the decametre regime with unprecedented resolution, opening new opportunities for studies of pulsars, exoplanetary systems, galaxy clusters, and other low-frequency phenomena.
}


   \maketitle

\nolinenumbers
\section{Introduction}

The decametre band ($\sim 10 - 30$ MHz) is the lowest-frequency window accessible to ground-based observatories and provides a unique view of several astrophysical and geophysical processes. Decametre observations are well suited to detecting aged, low-energy electron populations in extragalactic systems such as active galactic nuclei and galaxy clusters, because their synchrotron emission is less affected by radiative ageing than at higher radio frequencies \citep{2008A&ARv..15...67M,cassano2006statistics,mandal2020revived,shulevski2019first,tabatabaei2017radio}. These observations are also important for detecting emission produced by inefficient particle acceleration or re-acceleration mechanisms, where the synchrotron spectrum can steepen, curve, or become undetectable at higher frequencies \citep{de2017gentle,hodgson2021ultra}.

At the same time, spectra in the decametre regime can be shaped by turnover processes such as Razin suppression, synchrotron self-absorption, and free--free absorption \citep{condon1992radio,callingham2015broadband}. These processes often make sources fainter at the lowest frequencies, but they also provide diagnostics of the emitting plasma and surrounding medium, for example, in megahertz-peaked-spectrum sources.

This frequency band is also the natural regime for several coherent emission processes. In the Solar System, coherent solar bursts and Jupiter's intense auroral radio emission are prominent at these frequencies, with the latter observed only below 40 MHz \citep{zarka1998auroral,zarka2004fast,morosan2014lofar}. Analogous cyclotron emission from magnetised exoplanets is also expected in the decametre band \citep{zarka2007plasma,griessmeier2007predicting}, and detection of this emission would provide a direct measurement of exoplanetary magnetic field strengths. In addition, propagation through the terrestrial ionosphere leaves measurable imprints on the observed radio sky, making decametre observations a valuable probe of the ionosphere \citep{mevius2016probing,helmboldt2020ionospheric}. 

\begin{figure}
    \centering
    \includegraphics[width=1\linewidth]{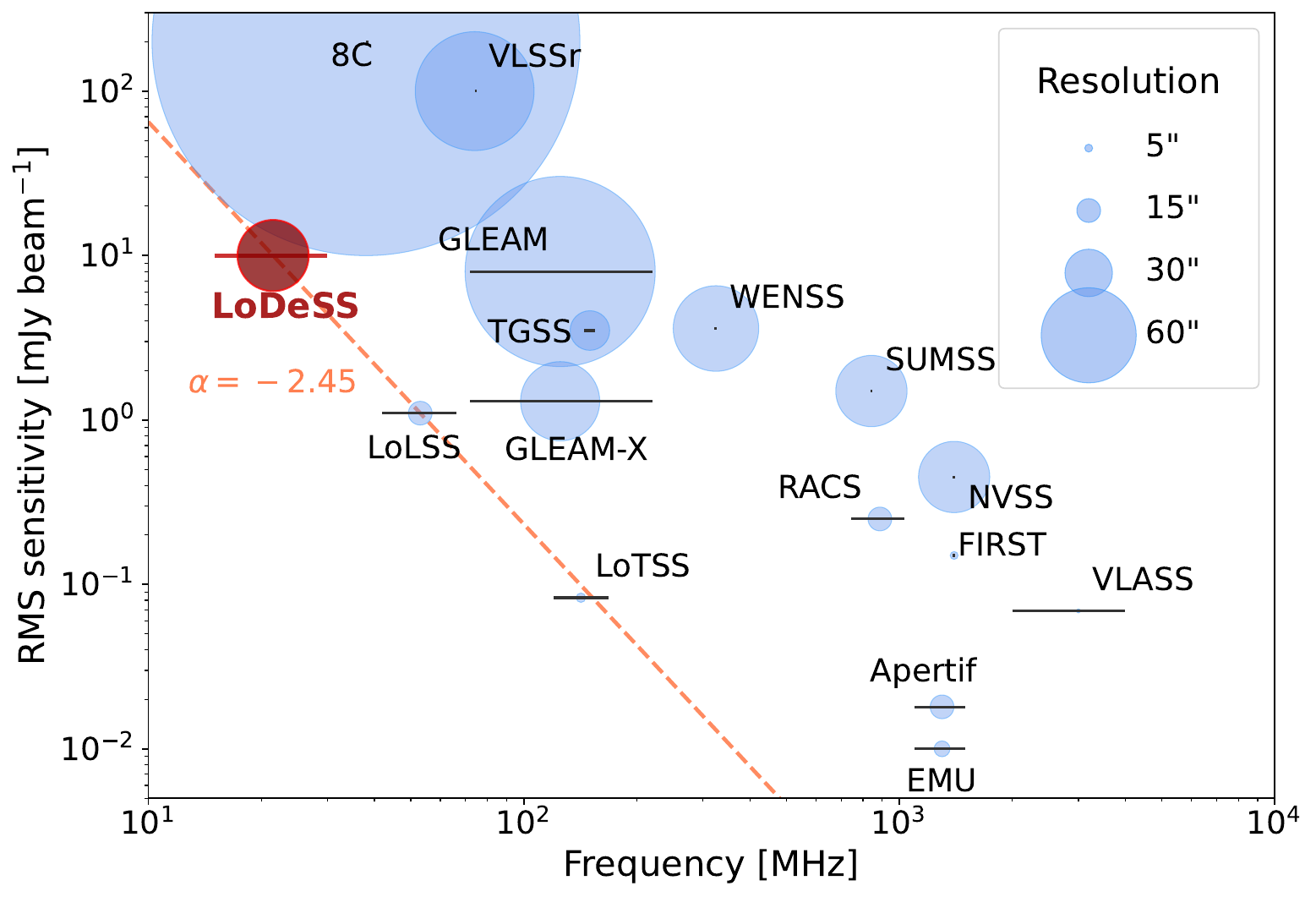}
    \caption{RMS sensitivity of radio surveys as a function of the observing frequency. Horizontal bars indicate the frequency coverage, while the circle sizes represent the angular resolution, as illustrated in the top-right legend. The red marker highlights the survey presented in this work. For surveys with very high angular resolution, the corresponding markers may be too small to be clearly visible at the scale of the figure.} The dotted line indicates the spectral index corresponding to the sensitivity difference between this work and the LOFAR LBA Sky Survey (LoLSS). References: 8C \citep{rees1990deep}, Galactic and Extragalactic All-sky Murchison Widefield Array Survey \citep[GLEAM - ][]{hurley2017galactic}, GLEAM Extended \citep[GLEAM-X - ][]{hurley2022galactic}, TIFR GMRT Sky Survey– Alternative Data Release 1 \citep[TGSS ADR1 - ][]{intema2009ionospheric}, VLA Low-frequency Sky Survey redux \citep[VLSSr - ][]{lane2014very}, Faint Images of the Radio Sky at Twenty Centimetres \citep[FIRST - ][]{becker1995first}, 1.4GHz NRAO VLA Sky Survey \citep[NVSS - ][]{1998AJ....115.1693C}, The Westerbork Northern Sky Survey \citep[WENSS - ][]{rengelink1997westerbork}, Sydney University Molonglo Sky Survey \citep[SUMSS - ][]{bock1999sumss}, Apertif \citep{adams2022first}, Evolutionary Map of the Universe \citep[EMU - ][]{norris2011emu}, VLA Sky Survey \citep[VLASS - ][]{lacy2020karl}, LOFAR Two-metre Sky Survey \citep[LoTSS - ][]{shimwell2026lofar}, LOFAR LBA Sky Survey \citep[LoLSS - ][]{de2021lofar}, Rapid ASKAP Continuum Survey \citep[RACS low - ][]{duchesne2025rapid}.
    \label{fig:intro}
\end{figure}

Existing wide-area radio surveys have covered a broad range of frequencies, from the gigahertz and decimetre-wave regimes (e.g. VLASS \citep{lacy2020karl}, EMU \citep{norris2011emu}) to the metre-wave regime (e.g. LoTSS \citep{shimwell2017lofar,shimwell2019lofar,shimwell2022lofar,shimwell2026lofar}, LoLSS \citep{de2021lofar}, GLEAM \citep{hurley2017galactic}). Recent advances in observing capabilities, computing power, and data processing techniques have facilitated wide-field surveys at an unprecedented depth and angular resolution. \autoref{fig:intro} shows the frequency coverage, sensitivity, and angular resolution of several representative wide-field radio surveys.

At decametre wavelengths, however, wide-area surveys remain much more limited. Earlier surveys included the Ukrainian T-shaped Radio telescope, second modification (UTR-2), the Clark Lake array, the Dominion Radio Astrophysical Observatory (DRAO) 22~MHz survey, and the Cambridge 8C survey, which provided important maps and source catalogues at frequencies of a few tens of megahertz \citep{braude1978decametric,kassim1988clark,roger1999radio,rees1990deep}. These efforts formed the basis of decametre radio astronomy for several decades, but their angular resolution and sensitivity were limited by the instrumental capabilities available at the time. More recently, preliminary studies over small areas of the sky have shown that sub-arcminute decametre imaging is feasible with the Low Frequency Array (LOFAR) Low Band Antenna (LBA) system \citep{osinga2024probing,groeneveld2024characterization,groeneveld2025serendipitous}. Nevertheless, a sensitive wide-area decametre survey at a comparable angular resolution is still lacking.

The decametre regime remains largely unexplored because observations at these frequencies are technically challenging. The terrestrial ionosphere introduces rapid temporal, spectral, and spatial phase variations across both the array and the wide field of view, making standard calibration practices insufficient \citep{de2019systematic}. These difficulties are amplified because the decametre band lies close to the ionospheric plasma resonance frequency ($\sim 1$--$10$~MHz), where standard high-frequency approximations for the ionospheric phase become increasingly inaccurate and higher-order terms are required \citep{de2018effect}. The effects are also strongly coupled to solar activity, so even modest solar disturbances can make observations difficult or impossible to calibrate \citep{mevius2016probing}. In addition, the large fields of view of low-frequency dipole arrays increase contamination from bright sources entering through sidelobes, while the high sky temperature and reduced sensitivity limit the signal-to-noise ratio (S/N) of calibration sources and therefore the number of directions that can be solved for. Finally, radio-frequency interference (RFI) is particularly severe in this regime, primarily due to shortwave radio transmissions below 30 MHz. Ionospheric reflection can propagate signals from distant and sub-horizon transmitters into the field of view, further increasing the interference \citep{offringa2013lofar, blaszkiewicz2021comparative}.

Modern instruments capable of sensitive imaging in the decametre regime remain relatively rare. One of the most sensitive facilities for exploring this frequency range is LOFAR \citep{VanHaarlem2013}, alongside instruments such as the Ukrainian T-shaped Radio Telescope \citep{braude1978decametric}, Clark Lake \citep{kassim1988clark}, Dominion Radio Astrophysical Observatory \citep{roger1999radio}, Owens Valley Long Wavelength Array \citep{eastwood201921}, and NenuFAR \citep{zarka2015nenufar}. Among these, LOFAR provides a unique combination of sensitivity and angular resolution owing to its large collecting area and long interferometric baselines. 

LOFAR is a European radio interferometer, with baselines of up to 3000~km. The highest concentration of LOFAR antenna fields is in the Netherlands, with baselines of up to 120~km. Each station hosts two antenna systems: the Low Band Antenna (LBA), operating at 10–90 MHz, and the High Band Antenna (HBA), covering 110–250 MHz. The LBA system consists of wide-field dipole antennas, whose field of view becomes $\sim 9^\circ$ in the decametre band. LOFAR LBA is therefore uniquely capable of performing wide-field, sub-arcminute imaging near the plasma cut-off frequency of the ionosphere.

The LOFAR Decametre Sky Survey (LoDeSS) is designed to explore the northern decametre radio sky with the LOFAR LBA system in the frequency range from 15 to 30~MHz (\autoref{fig:intro}). In this paper, we present the first data release of LoDeSS, including calibrated images and an associated source catalogue. Further details on the survey design, observing strategy, and data products are provided in Sect.~\ref{sec:survey}.

This paper is structured as follows. In Sect.~\ref{sec:survey} we describe the LOFAR Decametre Sky Survey, including the observing strategy, survey coverage, and observational setup. Sect.~\ref{sec:reduction} presents the data reduction and calibration procedures. The public data products released in this work are described in Sect.~\ref{sec:datarelease}, including the image products and the construction of the source catalogue. In Sect.~\ref{sec:results}, we evaluate the quality of the survey through measurements of the sensitivity, astrometric accuracy, flux density scale, and polarisation properties. Then Sect.~\ref{sec:discussion} discusses the challenges of calibrating and imaging in the decametre regime. Finally, Sect.~\ref{sec:conclusion} summarises the results and draws the main conclusions of this work.

\section{The LOFAR Decametre Sky Survey}\label{sec:survey}

\subsection{Survey overview}

LoDeSS (project code LT16\_014) is designed to map the northern sky above a declination of 20$^\circ$ between 15 and 30~MHz. The survey consists of 362~pointings, each with a total integration time of approximately five hours. Observations were carried out using a multi-beam strategy, with five target fields observed simultaneously, together with one calibrator beam. The total telescope time is 365~hours. Observations were carried out between June 2021 and June 2023, as solar activity rose from approximately 17\% to 78\% of the Solar Cycle 25 maximum. A summary of the survey parameters is provided in \autoref{tab:survey}.

\begin{table}[!htp]
    \centering
    \caption{LoDeSS observational setup}
    \begin{tabular}{lc}
    \hline \hline
      Pointings   &  362 \\ 
      Separation of pointings at 30~MHz   &  6.2$^\circ$ \\
      Integration time per pointing & 3-8\,h \\
      Frequency range & 15--30~MHz \\
      Angular resolution & $\sim45\arcsec$ \\
      RMS & 11.5~mJy~beam$^{-1}$ \\
      Time resolution & 1~s \\
      \hspace{1cm}-- After averaging & 4~s \\ 
      Frequency resolution &  3.052~kHz \\
      \hspace{1cm}-- After averaging & 24.414~kHz\\
    \hline
    \end{tabular}
    \label{tab:survey}
\end{table}

\subsection{Observational setup}

Observations were performed using the LOFAR Low Band Antenna (LBA) system with the receiver high-pass filter set to 10\,MHz. Hardware limitations require a selection of 48 out of the 96 LBA dipoles per station. The LBA\_SPARSE\_EVEN configuration was used, in which the selected dipoles are distributed across both the inner and outer parts of the station. This configuration provides a large primary beam, with a full width at half maximum (FWHM) of approximately 8.7$^\circ$ at 30\,MHz, increasing to 17.4$^\circ$ at 15\,MHz, and reduced sidelobe levels, making it well suited for wide-field decametre imaging.

The observations used up to 38 Dutch stations, with typically 34--38 available. International stations were excluded to reduce the calibration complexity on long baselines at these frequencies and to reduce the data volume \citep{morabito2016lofar,groeneveld2022pushing}.

Observations were performed in multi-beam mode, with six simultaneous beams: one targeting a primary calibrator and five on science fields. The primary calibrators used were 3C196 and 3C380, both $\sim70$ Jy at 159 MHz, because their spectra remain well characterised across the low-frequency band, unlike many standard calibrators that turn over in or near the decametre regime \citep{scaife2012broad}.

\subsection{Observing strategy}

Observations were conducted during nighttime to minimise radio-frequency interference and ionospheric disturbances \citep{groeneveld2024characterization}. The total integration time of five hours per pointing was obtained through multiple shorter observations, typically 60--120 minutes in duration, distributed over several days, weeks, or months. This strategy mitigates the impact of poor ionospheric conditions by increasing the likelihood of obtaining some usable data on each field.

The nominal integration time per pointing was 5\,h, although the effective integration time varied between fields. Observations were inspected after acquisition, and data affected by poor ionospheric conditions were re-observed when possible. In some cases, part of the original data were retained, resulting in total integration times exceeding 5\,h. In other cases, fields were not re-observed, and only the usable fraction is included, leading to integration times as low as 3\,h.

\subsection{Pointing layout}

\begin{figure}[!htp]
    \centering
    \includegraphics[width=0.85\linewidth]{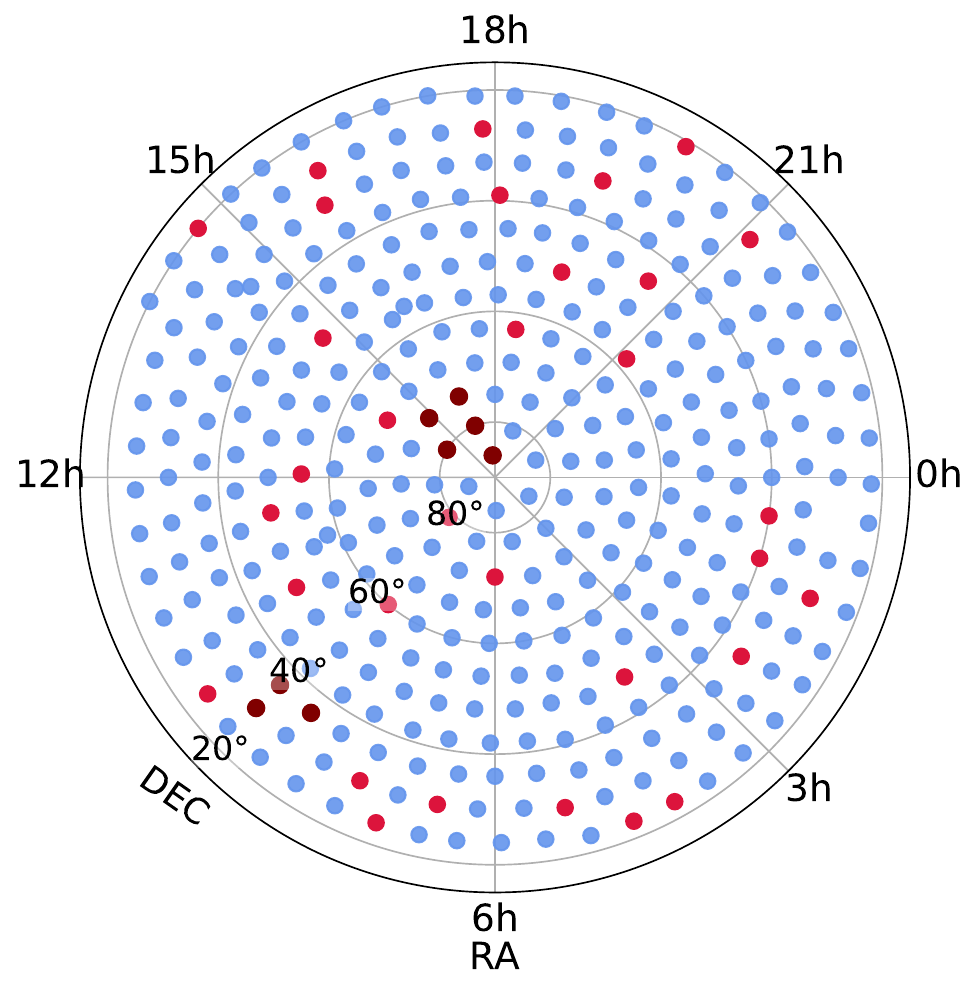}
    \caption{Sky coverage of LoDeSS. Each marker represents a survey pointing. The fields presented in this work are highlighted in red, with fields included in mosaics shown in dark red, while the remaining pointings are shown in blue.}
    \label{fig:pointings}
\end{figure}

The survey pointings are arranged in a spiral pattern to provide uniform sky coverage. Adjacent pointing centres are separated by $\mathrm{FWHM}/\sqrt{2}$ at the top of the band, corresponding to $\approx 6.2^\circ$, following the strategy adopted by the LOFAR Two-Metre Sky Survey \citep[LoTSS;][]{shimwell2017lofar}. The primary-beam FWHM at 30 MHz is approximately $8.7^\circ$, so this spacing provides sufficient overlap between adjacent fields ($\approx 2.5^\circ$ at 30~MHz). This improves sensitivity uniformity and makes flux-density measurements more robust across the survey area. The full pointing layout is shown in Fig.~\ref{fig:pointings}.

\subsection{Survey status}

All 362 planned pointings have been observed. However, not all datasets meet the quality requirements for calibration and imaging. Data quality is primarily limited by ionospheric conditions and radio-frequency interference, and not all affected observations could be repeated due to the time constraints of the project.

We processed 44 fields, of which 40 were successfully calibrated and included in this first data release. The remaining four were excluded due to severe interference or calibration failures caused by poor ionospheric conditions. The released fields were selected based on known exoplanet candidates, resulting in a non-contiguous sky distribution. Of the 40 released fields, 14 (35\%) are adjacent to at least one other released field and were included in mosaics, while the remaining 26 (65\%) are spatially disconnected. We find no evidence for radio emission from these candidates, and the corresponding upper limits will be presented in a forthcoming paper (Cordun et al., in prep.).

The quality of the processed fields varies with observing conditions and field characteristics, with typical noise levels of 10--20 mJy\,beam$^{-1}$. Further improvements in sensitivity and uniformity are expected through mosaicing of all fields in future releases. The remaining survey fields are currently being processed and will be included in a subsequent data release.

\section{Data reduction}\label{sec:reduction}

The data were reduced with the Library for Low-Frequency \citep[LiLF;][]{de2019systematic}, the calibration pipeline developed for LOFAR LBA. Calibration was performed with DP3 \citep{van2018astrophysics}, imaging with WSClean \citep{offringa2014wsclean, offringa2017optimized}, solution inspection with LoSoTo \citep{de2019systematic}, and data compression with Dysco \citep{offringa2016compression}. The calibration strategy consists of pre-processing, primary-calibrator calibration, transfer of these solutions to the target fields, and direction-dependent self-calibration. For the decametre data presented here, several modifications to the standard processing of LBA data above 30\,MHz \citep{de2019systematic} were required to account for stronger ionospheric effects and radio-frequency interference, including additional flagging to mitigate artefacts. The total fraction of flagged data varies between 7\% and 35\% across fields. The flagged data are not uniformly distributed across frequency: the 20--26~MHz range is often most strongly affected by an intermittent interference source whose origin has not yet been identified. The impact of this flagging on image sensitivity is described in Sect.~\ref{sec:results_1}. The pipeline is available on GitHub\footnote{\url{https://github.com/revoltek/LiLF}, commit \texttt{d5f4104}}.

\subsection{Pre-processing and primary calibration}

During pre-processing, the data are averaged to 24.4\,kHz in frequency and 4\,s in time. In the primary-calibrator pipeline, the individual measurement sets, one per subband, are first merged in frequency, and the bandpass is flattened using a precomputed correction from \citet{VanHaarlem2013}. Radio-frequency interference is then flagged with AOFlagger \citep{offringa2012morphological}, primarily targeting strong broadband RFI below 30\,MHz. On average, this removes 20\% of the data. The calibrator solutions, namely polarisation misalignment, phase solutions including clock and ionospheric effects, and bandpass corrections, are then derived following the standard LiLF procedure. These solutions are transferred directly to the target before the measurement sets are combined and split into 60-minute blocks for further processing.

\subsection{Self-calibration} \label{sec:selfcal}

\begin{figure*}[!htp]
    \centering
    \includegraphics[width=0.9\linewidth]{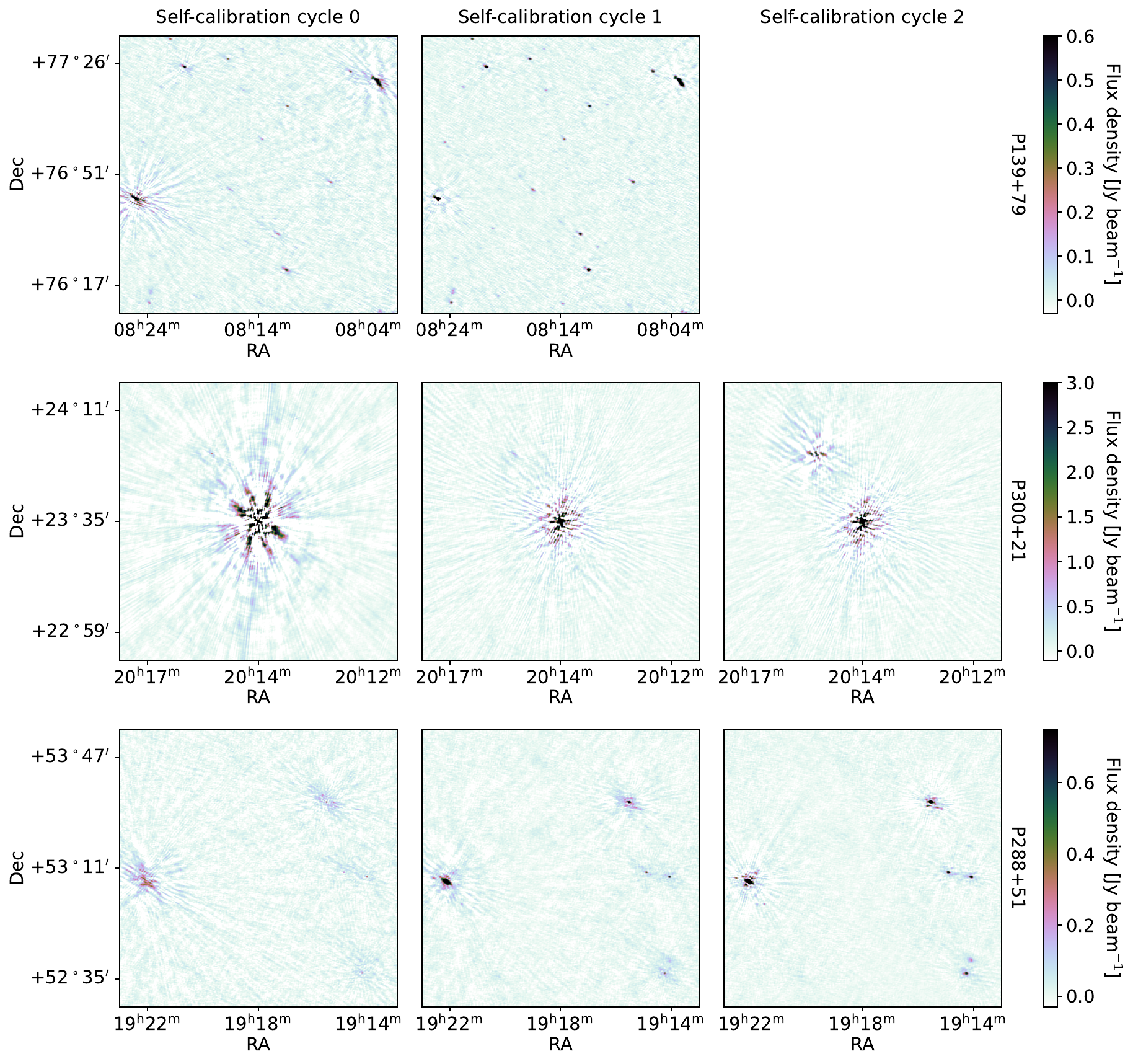}
    \caption{Progressive improvement during direction-dependent phase-only self-calibration for three representative LoDeSS fields. The columns show successive self-calibration cycles, and rows correspond to pointings P139+79 (good), P300+21 (poor), and P288+51 (mediocre). The standard field P139+79 required two self-calibration cycles, while the bright-source fields P300+21 and P288+51 required three. For the bright-source fields, we also applied the additional direction-independent amplitude calibration described in Sect.~\ref{sec:selfcal}, with amplitudes normalised to preserve the flux-density scale. In the first cycle of P288+51, the single calibration direction lies far from the region shown, resulting in strong residual phase errors and ionospheric decorrelation that suppress the recovered source flux densities.}
    \label{fig:comp}
\end{figure*}

To correct for the strong spatial variation of the ionosphere across the field of view, LiLF performs direction-dependent calibration of the target field, without an initial direction-independent self-calibration step. First, the ionospheric differential Faraday rotation is corrected, and initial direction-dependent phase solutions are derived against an external sky model. When available, this model is taken from LoTSS DR3. Otherwise, we use a combination of WENSS \citep{rengelink1997westerbork}, NVSS \citep{1998AJ....115.1693C}, VLSSr \citep{lane2014very}, and TGSS \citep{intema2017gmrt}. The calibration facets are defined from this initial sky model in the following way. Sky-model components within $2\arcmin$  are first grouped into source patches, with patch positions defined by their apparent-flux-weighted mean coordinates. Nearby bright patches within $1\arcmin$ are then merged when their apparent flux density exceeds $0.5$ Jy. Each remaining patch above the cycle-dependent apparent-flux threshold is used as a facet centre. If this produces more than the allowed maximum of 10 facets in the first cycle or 22 facets in the second cycle, the flux threshold is increased so that only the brightest patches are retained. All other patches are assigned to the nearest facet centre using the Voronoi grouping scheme implemented in \texttt{LSMTool} \citep{lsmtool}. The direction-dependent phase solutions are then applied during imaging to produce an updated image, which is used to construct the sky model for the next calibration step. The facet with the highest flux density is selected as the subfield, and emission outside this subfield is subtracted from the visibilities. This isolates a high-signal-to-noise direction in which a stable phase-only solution can be obtained without contamination from the rest of the field. A direction-independent phase correction is then derived for this subfield, using the updated sky model, and applied to the full field. In the next cycle, direction-dependent phase solutions are solved for each calibration facet on top of these subfield phase corrections. This procedure typically allows 16--22 facets to be calibrated for each target field. The progressive improvement obtained for a representative standard field is shown in the top row of \autoref{fig:comp}.

Compared to the default implementation, we modified only the frequency constraint used by the DP3 solver, reducing the frequency range over which both the direction-dependent and subfield phase solutions are constrained by one-third. This allows the calibration to better follow the stronger frequency dependence of ionospheric phase variations at decametre wavelengths. We used a frequency smoothness constraint ranging from $\sim$1\,MHz for the outer remote stations to $\sim$5\,MHz for the core stations. The solution time intervals were left unchanged from the default pipeline settings: 15\,min for the core stations and 1--4\,min for the remote stations, with the shortest intervals used for the outermost stations.

For fields without a source brighter than 100 Jy, we followed the target-field calibration strategy described above. Fields dominated by bright sources require a modified strategy, because full-field solutions are degraded by direction-dependent effects and early image-based clean-component models are typically dominated by artefacts. For these fields, the initial direction-dependent calibration cycle is therefore restricted to a single facet containing the dominant bright source. The calibration then proceeds as described above until the subfield-calibration stage, where we derive the differential Faraday rotation correction on this subfield and perform two phase-only self-calibration cycles followed by amplitude calibration. This mitigates strong ionospheric scintillation effects, which would otherwise lead to significant artefacts and divergence of the imaging process \citep{fallows2020lofar}. Although shorter time and bandwidth solution intervals would be preferable, the S/N is insufficient to support them, and additional self-calibration cycles did not improve the results. The resulting amplitude solutions are normalised to preserve the flux density scale. After completing this additional calibration of the bright-source subfield, the pipeline returns to full-field direction-dependent calibration. The first full-field cycle is performed with a reduced number of facets and uses the external sky model, because an image-based model is not yet reliable in the presence of residual bright-source artefacts. Once this cycle has improved the resulting image, the following cycles proceed in the same way as for fields without a source brighter than 100 Jy, using image-based clean-component models. The progressive improvement produced by the self-calibration is illustrated in the middle and bottom panels of \autoref{fig:comp}.

All selected calibration facets produced usable phase solutions in the processed fields, including those observed under relatively poor ionospheric conditions. We therefore did not need to discard failed facets or transfer solutions from neighbouring calibration directions. This stability is aided by solving only for phase in the facet-based calibration, which is less susceptible to unstable solutions than simultaneous amplitude-and-phase calibration. Normalised amplitude solutions are derived only in a direction-independent fashion, and values outside the range 0.1--3 are flagged before application to prevent unstable or unphysical corrections.

\subsection{Post-processing}\label{sect:pp}

After imaging, we applied a global astrometric correction per field to align the source positions with the LoTSS DR3 catalogue \citep{shimwell2026lofar}. To ensure reliable cross-matching, we used unresolved sources following the criterion described in Sect.~\ref{sec:results_0}. In addition, we required source isolation: LoDeSS sources were required to have no neighbours within 90\arcsec, and LoTSS sources within 30\arcsec. With the unresolved and isolated criteria, the catalogues were cross-matched using the nearest counterparts, and a single global offset was derived for each field. The resulting offsets were small, below 2\arcsec\ in all fields.

\section{Data products}\label{sec:datarelease}
Our first data release comprises observations of 40 fields selected around well-known radio-emitting exoplanet candidates (Cordun et al., in prep.). Each field covers an effective area of approximately $8.7 \times 8.7$ deg$^2$. Due to partial overlap between neighbouring fields, the total sky coverage is $\sim 2500$ deg$^2$, corresponding to roughly 15\% of the full survey area. A summary of the individual fields and their observational properties is provided in Appendix~\ref{sec:ap1}. The final products are available at  LOFAR Surveys\footnote{\url{https://lofar-surveys.org/lodess.html}}.

\subsection{Sky images}

\begin{figure*}[!htp]
    \centering
    \includegraphics[width=\textwidth]{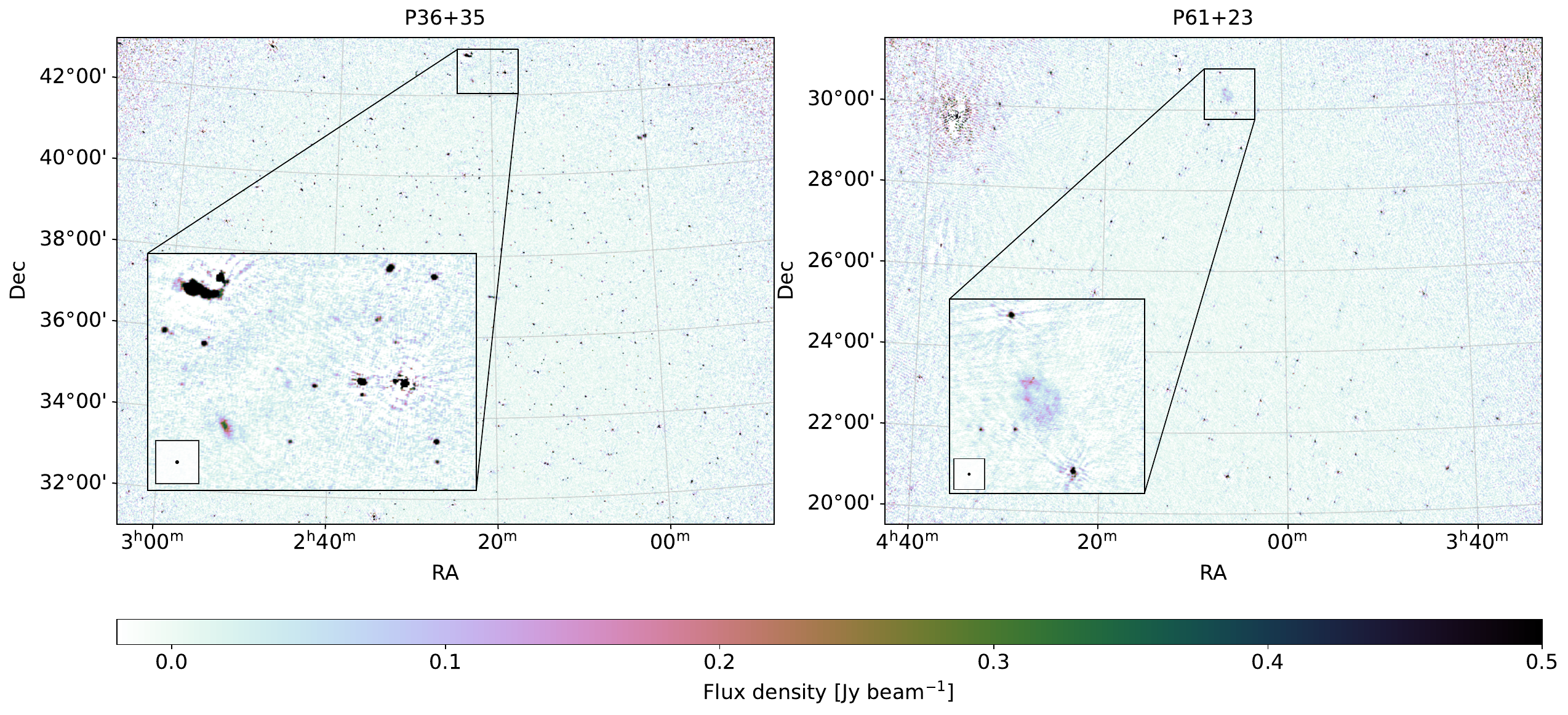}

    \caption{Representative LoDeSS images. The left panel shows the high-quality pointing P36+35 (5-h integration, rms 11.06~mJy~beam$^{-1}$). The inset contains 3C66B (top left), NGC891 (bottom left), and the variable pulsar PSR~J0218+4232 (right), whose variability causes residual deconvolution artefacts \citep{kumamoto2021flux}. The right panel shows the moderate-quality pointing P61+23 (5.5-h integration, rms 14.87~mJy~beam$^{-1}$), affected by ionospheric conditions and 3C123. The inset shows the galaxy group Nest200047. Noise increases towards the image edges because of primary-beam effects.}

    \label{fig:representative_images}
\end{figure*}

Because DR1 covers a non-uniform subset of the full survey, the fields span a wide range of observing conditions, including bright-source contamination, a range of declinations, and varying ionospheric conditions. For each field, we generated primary beam-corrected images using the full 15--30\,MHz bandwidth, corresponding to a central frequency of 23\,MHz. We convolved the images to a common circular restoring beam of 45\arcsec, consistent with the native angular resolution of the survey \citep{groeneveld2024characterization}. Representative examples are shown in \autoref{fig:representative_images}. PSR~J0218+4232 in the left panel exhibits flux-density variability during the observation. Pulsar flux densities can vary because of interstellar scintillation \citep{kumamoto2021flux}. Since synthesis imaging and deconvolution assume an approximately time-invariant sky, this variability produces a mismatch with the static source model and leaves local residual artefacts around the pulsar.

The rms noise is $\sim 12$ mJy beam$^{-1}$ at the field centre. While the theoretical sensitivity is approximately 10 mJy beam$^{-1}$, the achieved noise levels are typically higher (12--20 mJy beam$^{-1}$) due to residual ionospheric effects and data flagging. Moreover, the noise increases towards the image edges due to primary beam effects. These variations in image quality are quantified in Sect.~\ref{sec:results}.

\subsection{Mosaiced images}

We combined overlapping fields using the mosaicing procedure adopted for LoLSS DR1 \citep{de2021lofar}\footnote{Script at \url{https://github.com/revoltek/scripts/blob/master/mosaic.py}}. Before combination, the images were reprojected onto a common astrometric grid and convolved to a common circular restoring beam using the \texttt{--beamarm} and \texttt{--beamcirc} options of the aforementioned mosaicing script. Although the individual LoDeSS images already have circular restoring beams, this step ensures that the resolution is identical within the overlapping regions. Pixels below a primary-beam response of 0.3 were excluded.

\begin{figure*}
    \centering
    \includegraphics[width=\textwidth]{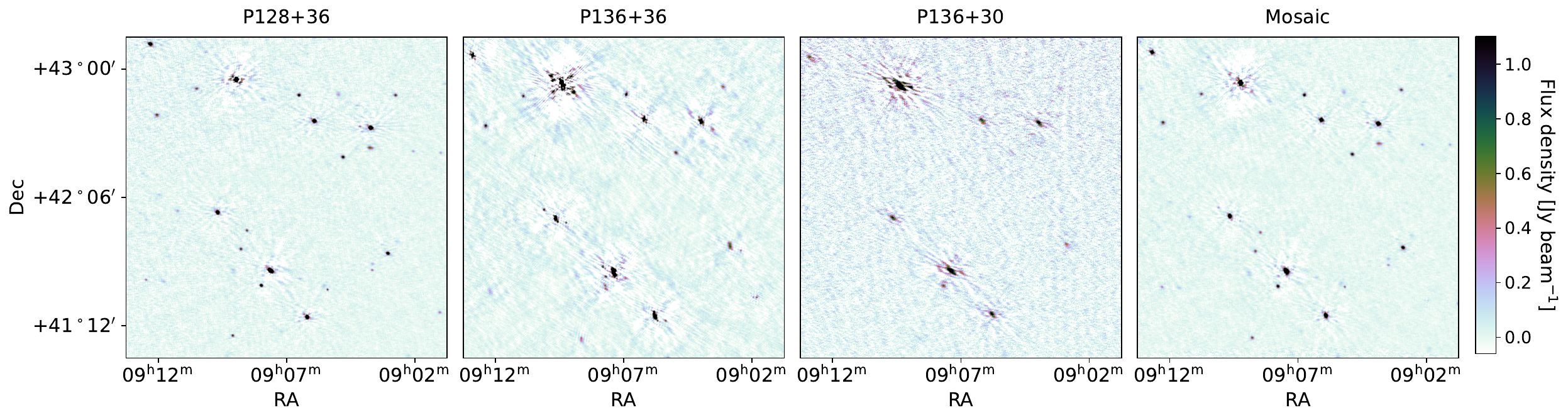}
    \caption{Example of a LoDeSS mosaiced subfield, shown alongside the three contributing overlapping fields: P128+36, P136+36, and P136+30. Their rms noise levels are 7.76, 15.36, and 9.46~mJy~beam$^{-1}$, respectively, while the resulting mosaic has an rms noise of 11.13~mJy~beam$^{-1}$. The mosaicing therefore primarily provides more uniform noise properties and improves the flux-density consistency, rather than uniformly reducing the rms noise relative to the individual fields.}
    \label{fig:mosaic_example}
\end{figure*}

The mosaic was formed as a weighted mean of the contributing images. The weights include the primary-beam response, the central image noise, and an additional field-dependent scale factor. This scale factor reduces the contribution of fields affected by poor ionospheric conditions or substantial flux-density loss, thereby reducing the flux-density scatter in overlapping regions (see Sect.~\ref{sec:results_3}). \autoref{fig:mosaic_example} shows three overlapping fields and the resulting mosaic. For this initial release, the scale factors were selected empirically on a field-by-field basis, based on the image quality and measured flux-density loss. The factors were adjusted iteratively to reduce the contribution of fields with poorer calibration quality in the overlap regions. An automated procedure will be developed for the final survey release.

\subsection{Source catalogues}\label{sec:source_catalogue}

We used \texttt{PyBDSF} \citep{mohan2015astrophysics} to construct source catalogues from our primary-beam-corrected images. Source islands were detected in the non-primary-beam-corrected images, where the noise is not amplified by division through the time-averaged primary-beam model. Source properties, including flux densities, were then measured from the corresponding primary-beam-corrected images.

We used detection thresholds of 5$\sigma$ for islands and 7$\sigma$ for peak pixels, which provided a good balance between sensitivity and limiting false detections around bright sources. The background noise was estimated from a spatially varying rms map using an adaptive rms box of 150 by 150 pixels and a sampling step of 15 pixels. In regions with S/N ratios higher than 50, we used a smaller box of 60 by 60 pixels to better capture local noise variations near bright sources. PyBDSF produces both a source catalogue and a Gaussian component catalogue. We did not group sources at the island level (\texttt{group\_by\_isl=False}), such that emission is not merged into a single source per island and complex emission may be represented by multiple source entries. This choice is motivated by residual ionospheric artefacts at these frequencies, which can make island grouping unreliable. For similar reasons, we disabled wavelet-based decomposition, which gave less robust catalogues for these images. We therefore performed source extraction using the standard PyBDSF island-based Gaussian fitting.

The structure of the source catalogue provided in this data release is described in Appendix~\ref{sec:ap2}. As the current release includes overlapping fields, some sources are detected multiple times. These duplicates were removed from the source catalogue by cross-matching sources within one synthesised beam and retaining the detection with the highest total flux density, to minimise flux density losses due to ionospheric decorrelation (see Sect.~\ref{sec:results_3}). Duplicate removal is applied only to the source catalogue -- the Gaussian component catalogue is left unchanged, as individual Gaussian components cannot be reliably associated across overlapping fields.

We classified the fields as good, mediocre, or poor based on their flux-density reliability. For each field, we calculated the median ratio between the flux density predicted at 23~MHz from the NVSS, LoTSS, and VLSSr measurements and the measured LoDeSS flux density, together with its semi-interquartile range (SIQR). As a general criterion, fields with a median ratio below 2.5 and an SIQR below 1.0 were classified as good, while fields with a median ratio of at least 4.5 or an SIQR of at least 2.0 were classified as poor. The remaining fields were classified as mediocre, with borderline cases additionally assessed based on the calibration solutions and image quality.

Moreover, while we release the images for fields with large flux-density uncertainties, which we classify as poor, we exclude them from the final source catalogue to avoid misleading potential users (see Appendix~\ref{sec:ap1}). For the remainder of the analysis, however, we retain all fields, including those excluded from the final source catalogue.

\subsection{Completeness and false positives}\label{sec:datarelease_3}

We assessed the reliability and completeness of the source catalogue following standard procedures adopted in radio surveys \citep[e.g.][]{heald2015lofar, de2023lofar}. The assessment used non-primary-beam-corrected images, as source detection was carried out on these data, while primary-beam-corrected images are used only for accurate flux density measurements.

\begin{figure}[!htp]
    \centering
    \includegraphics[width=0.9\linewidth]{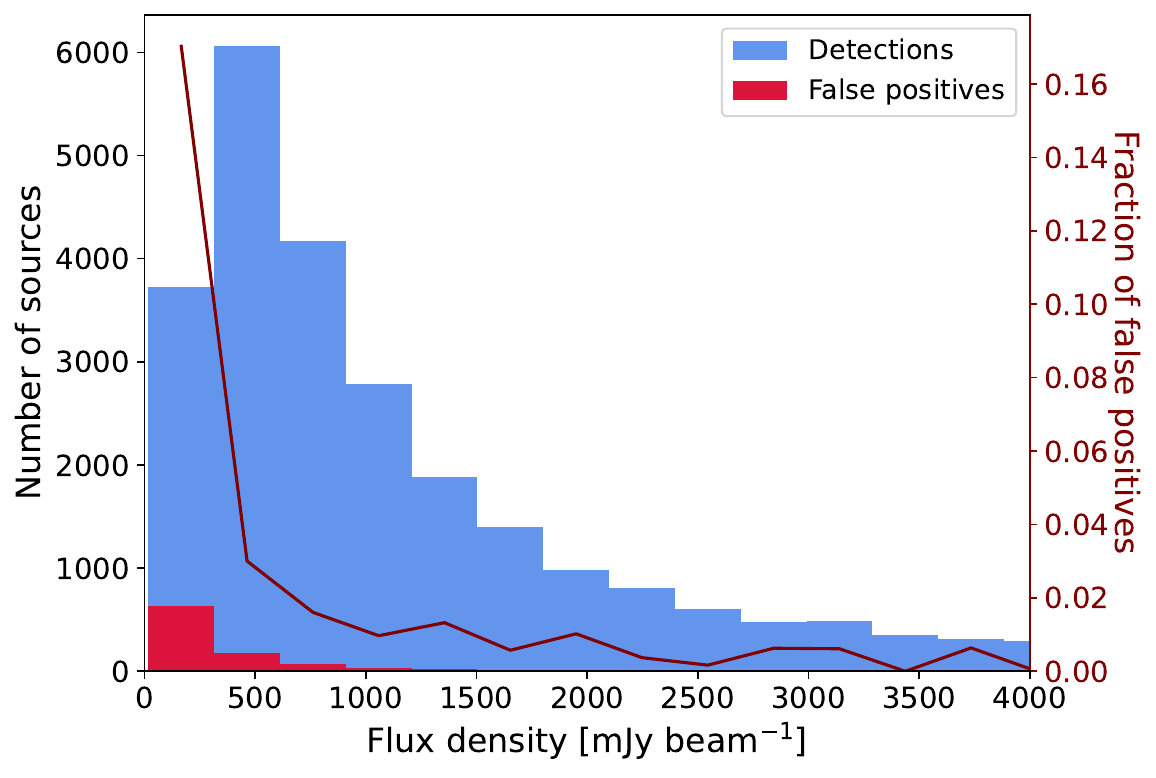}
    \caption{Distribution of false positives (red) and detected sources in LoDeSS (blue). The dark red line shows the fraction of false positives in each flux-density bin. The number of detections decreases in the faintest bin because elevated noise due to data flagging and shorter integration times reduces the survey sensitivity at these flux densities.}
    \label{fig:false_pos}
\end{figure}

We estimated the fraction of false detections by applying the same source-extraction procedure to sign-inverted \texttt{PyBDSF} residual images. Since these images do not contain real astrophysical sources, any detected components arise from noise fluctuations or residual calibration artefacts. To ensure consistency, we reused the rms map derived from the original images during source extraction. We find a spurious detection fraction of approximately 3.57\% (see \autoref{fig:false_pos}), with most false positives occurring at low flux densities and near bright sources, consistent with expectations for low-frequency radio surveys affected by residual calibration artefacts. The reduced number of detections in the lowest flux density bin is primarily driven by elevated noise levels from data flagging and shorter integration times.

We evaluated the survey completeness using injection--recovery tests on the \texttt{PyBDSF} residual images. These images contain the thermal and systematic noise that remain after source extraction and therefore provide a realistic environment for testing source recovery. We injected $10^4$ artificial point-like sources at random positions across each image, with flux densities drawn from a power-law source-count distribution of $dN/dS \propto S^{-1.6}$ \citep{wilman2008semi}. The injected sources were modelled as unresolved sources with a Gaussian profile matching the synthesised beam. Their flux densities ranged from twice the theoretical thermal noise limit, corresponding to 20~mJy beam$^{-1}$ for a 5-hour observation with no flagged data, to 3~Jy. The minimum flux density was scaled for each field according to its effective integration time, accounting for both observing duration and the fraction of flagged data. Source extraction was then repeated using the same parameters as for the real data. We additionally repeated the experiment using circular Gaussian sources with observed FWHM values of 1, 2, and 4 times the synthesised beam. For each size and field, we injected 2000 sources in batches of 500. The lower number of injected sources reflects the lower expected surface density of diffuse objects and limits artificial blending between them, whereas blending is intentionally retained in the point-source experiment to represent the much higher density of compact sources.

For the point-source experiment, a source was considered recovered if it was detected within three times the synthesised-beam FWHM of its input position and if the difference between the recovered and injected flux densities was within ten times the uncertainty on the recovered flux density, following a similar approach to that of \citet{de2023lofar}. The same positional and flux-consistency criteria were used for the resolved-source experiment. Since resolved sources may be represented by multiple \texttt{PyBDSF} catalogue entries, their recovered integrated flux density was calculated by summing all entries within one injected-source FWHM, with the corresponding uncertainties added in quadrature.

\begin{figure}[!htp]
    \centering
    \includegraphics[width=\linewidth]{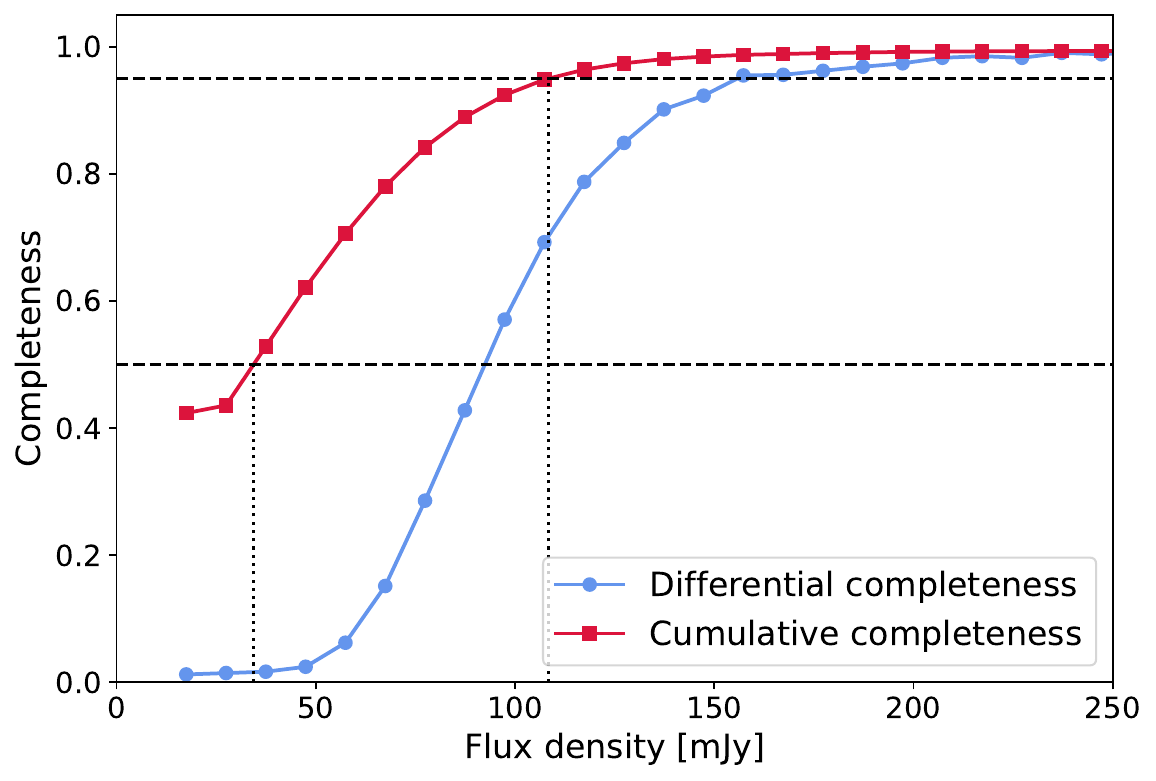}
    \caption{Completeness of the LoDeSS survey for point-like sources. The blue curve shows the differential completeness (fraction of detected sources as a function of flux density), while the red curve shows the cumulative completeness (fraction above a given flux density). The dashed lines indicate the 50\% and 95\% completeness levels at 34.4 and 108.4 mJy~beam$^{-1}$, respectively.}
    \label{fig:completeness}
\end{figure}

\begin{figure}[!htp]
    \centering
    \includegraphics[width=\linewidth]{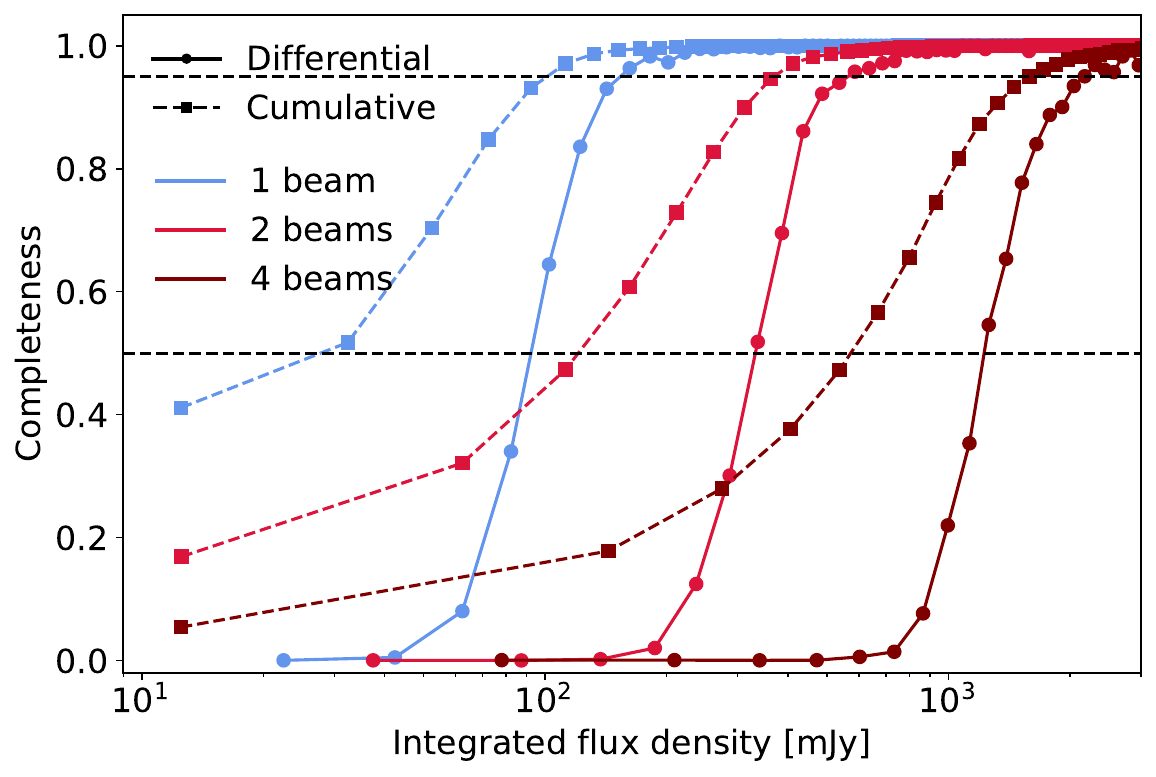}
    \caption{Completeness of LoDeSS for circular Gaussian sources with observed FWHM values of 1, 2, and 4 times the synthesised beam. Colours distinguish the three source sizes, while solid and dashed curves show the differential and cumulative completeness, respectively. The one-beam curve provides point-source control, whereas the two- and four-beam curves show reduced recovery for increasingly resolved sources.}
    \label{fig:diffuse_completeness}
\end{figure}

By applying these procedures to each field, we derived the recovery fraction as a function of flux density and source size. The point-source and resolved-source results are shown in \autoref{fig:completeness} and \autoref{fig:diffuse_completeness}, respectively, with the corresponding 50\% and 95\% differential and cumulative completeness limits summarised in \autoref{tab:completeness_limits}. The one-beam injections provide a point-source control and agree with the independent point-source completeness experiment. Completeness decreases with increasing source size because the flux of resolved sources is distributed over a larger area, lowering their peak surface brightness and making them more likely to fall below the local noise and detection threshold.

\begin{table}[!htp]
    \centering
    \caption{Differential and cumulative completeness limits for point-like and resolved sources.}
    \label{tab:completeness_limits}
    \begin{tabular}{lcccc}
        \hline
        \hline
        Source model
        & \multicolumn{2}{c}{Differential }
        & \multicolumn{2}{c}{Cumulative } \\
        & \multicolumn{2}{c}{completeness }
        & \multicolumn{2}{c}{completeness } \\
        & \multicolumn{2}{c}{ [mJy]}
        & \multicolumn{2}{c}{ [mJy]} \\
        & 50\% & 95\% & 50\% & 95\% \\
        \hline
        Point source      & 92.4 & 155.7 & 34.4 & 108.4 \\
        Gaussian, 1 beam  & 93.0  & 154.3   & 28.8  & 104.8   \\
        Gaussian, 2 beams & 332.5  & 564.5   & 122.3  & 370.1   \\
        Gaussian, 4 beams & 1128.9  & 2161.5   & 575.7  & 1594.3   \\
        \hline
    \end{tabular}
\end{table}

\section{Data quality} \label{sec:results}

In this section, we assess the overall data quality of the release. We quantify the accuracy and precision of the flux-density scale, source size measurements, and astrometry, and also evaluate the image noise and Stokes~V leakage. Data affected by stronger ionospheric disturbances are more challenging to calibrate and tend to show more artefacts, higher noise levels, and increased source smearing. Unless stated otherwise, all quality assessments presented in this section are based on the individual primary-beam-corrected field images. The flux-density analysis in Sect.~\ref{sec:results_3} also compares measurements from individual fields with their corresponding mosaics.

\subsection{Source sizes}\label{sec:results_0}

Separating point-like from extended sources is difficult at these frequencies because residual ionospheric phase errors and calibration imperfections broaden the sources, so even intrinsically unresolved sources often have integrated flux densities larger than their peak flux densities \citep{de2021lofar}. Following \citet{shimwell2022lofar}, we characterise this effect using $R = \ln(S_I / S_P)$, where $S_I$ is the integrated flux density (\texttt{Total\_flux} in the PyBDSF catalogue) and $S_P$ is the peak flux density (\texttt{Peak\_flux}). We then study how $R$ changes with S/N, defined as $S/N = S_P / \sigma_{S_P}$, where $\sigma_{S_P}$ is the local noise (\texttt{Isl\_rms}).

We do not rely on the \texttt{PyBDSF} single-component classification in our catalogue alone, because residual ionospheric smearing can broaden intrinsically compact sources and therefore affect this classification. Instead, we construct a cleaner compact-source sample by cross-matching our catalogue with LoTSS DR3 \citep{shimwell2026lofar}.

We first select isolated, high-signal-to-noise sources in both catalogues. In our catalogue, sources are required to have no neighbouring source within $90\arcsec$, corresponding to approximately twice the synthesised beam. For LoTSS, we adopt an isolation radius of $30\arcsec$. In both catalogues, we require a minimum S/N of 10. For the LoTSS reference sample, we additionally retain only sources consistent with being unresolved according to the LoTSS DR2 \(99.9\%\) envelope \citep{shimwell2022lofar},
\begin{equation}
    R_{99.9}(S/N) = 0.42 + \frac{1.08} {1 + \left( \frac{S/N}{96.57} \right)^{2.49}}.
\end{equation}

Sources were matched to their nearest LoTSS counterpart within a radius of $60\arcsec$, retaining only one-to-one associations. Given the strict selection criteria, in particular the isolation requirement, this matching radius is expected to minimise false associations while maintaining high completeness.

\begin{figure*}[!htp]
    \centering
    \includegraphics[width=\linewidth]{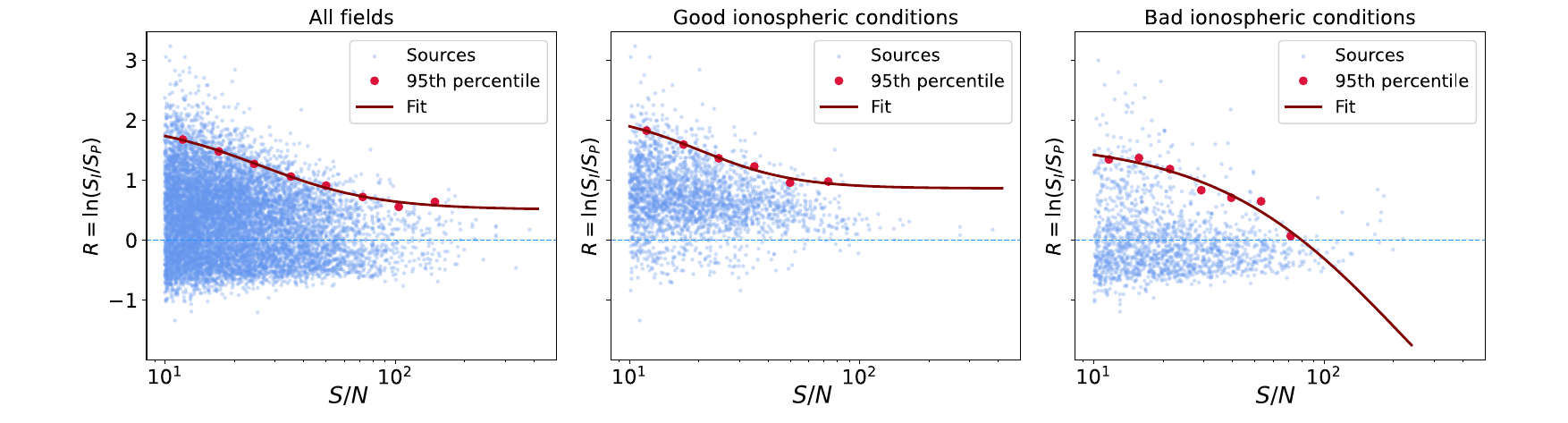}
    \caption{$R = \ln(S_I/S_P)$ as a function of the S/N for sources that are matched to isolated sources in LoTSS and have $S/N \geq 10$. The fitted envelope is overlaid in dark red. We show the distribution for all fields (left), fields with good ionospheric conditions (middle), and fields with the poorest ionospheric conditions (right).}
    \label{fig:sources}
\end{figure*}

At these decametre frequencies, residual ionospheric effects still smear many sources. As a result, the distribution of $R$ is generally shifted towards values above zero, as shown in the left panel of \autoref{fig:sources}. At high $S/N$, the values of $R$ form a relatively narrow sequence, whereas the scatter increases significantly towards low $S/N$. The distribution also depends on the observing conditions. In fields with comparatively good ionospheric conditions, shown in the middle panel of \autoref{fig:sources}, the positive offset in $R$ is consistent with residual source broadening, while the sources remain concentrated around a relatively well-defined sequence. Fields affected by poor ionospheric conditions, shown in the right panel, exhibit a substantially larger dispersion and a different dependence on $S/N$. In these fields, many sources have $R<0$, corresponding to $S_I/S_P<1$. This effect is likely caused by ionospheric decorrelation and the resulting loss of recovered source flux density. In particular, imperfect recovery of the broadened, low-surface-brightness emission can lead to a stronger underestimation of the integrated flux density than of the fitted peak flux density. This behaviour is consistent with the larger flux-density losses measured in fields affected by poor ionospheric conditions, as discussed in Sect.~\ref{sec:results_3}.

Despite these differences, the upper boundary of the combined distribution is predominantly determined by sources from the better-quality fields, while most sources from poor-quality fields lie below it. We therefore retain an envelope derived from the full LoDeSS-DR1 sample as a uniform survey-level classification criterion. This choice is conservative against falsely classifying compact sources as extended, although the completeness of the extended-source classification may be reduced in fields affected by poor ionospheric conditions. 

We therefore define an empirical upper envelope directly from the LoDeSS-DR1 catalogue. The data are binned in logarithmic intervals of $S/N$. In each bin, we measure the 95th percentile of the $R$ distribution. We use ten bins and require at least 25 sources per bin, so that poorly populated bins do not affect the fit. The resulting percentile values are then fitted with a smooth function of the form
\begin{equation}
R_{\mathrm{env}}(\mathrm{S/N}) =
a + \frac{b}{1 + (\mathrm{S/N}/c)^d},
\end{equation}
where $a$, $b$, $c$, and $d$ are free parameters. The function was fitted to the binned 95th-percentile values using nonlinear least squares implemented with \texttt{scipy.optimize.curve\_fit}. As no parameter bounds were
specified, the Levenberg--Marquardt algorithm was used.

The best-fitting envelope is shown in the left panel of \autoref{fig:sources}
and is given by

\begin{equation}
    R_{\mathrm{env}}(S/N) =
    0.51 + \frac{1.47}{1 + \left( \frac{S/N}{25.81} \right)^{1.69}}.
\end{equation}

We classify sources as point-like when $R \leq R_{\mathrm{env}}(\mathrm{S/N})$, and as extended otherwise. This provides a signal-to-noise-dependent separation that is better suited to the data than a fixed threshold in $S_I/S_P$. The envelope is derived from the combined sample rather than separately for each pointing. Since ionospheric smearing can vary between pointings, this approach may introduce field-dependent biases, and the resulting classification should be interpreted as an empirical survey-level point-like and extended separation.

\subsection{Sensitivity}\label{sec:results_1}

\begin{figure*}[!hbt]
    \centering
    \includegraphics[width=0.9\linewidth]{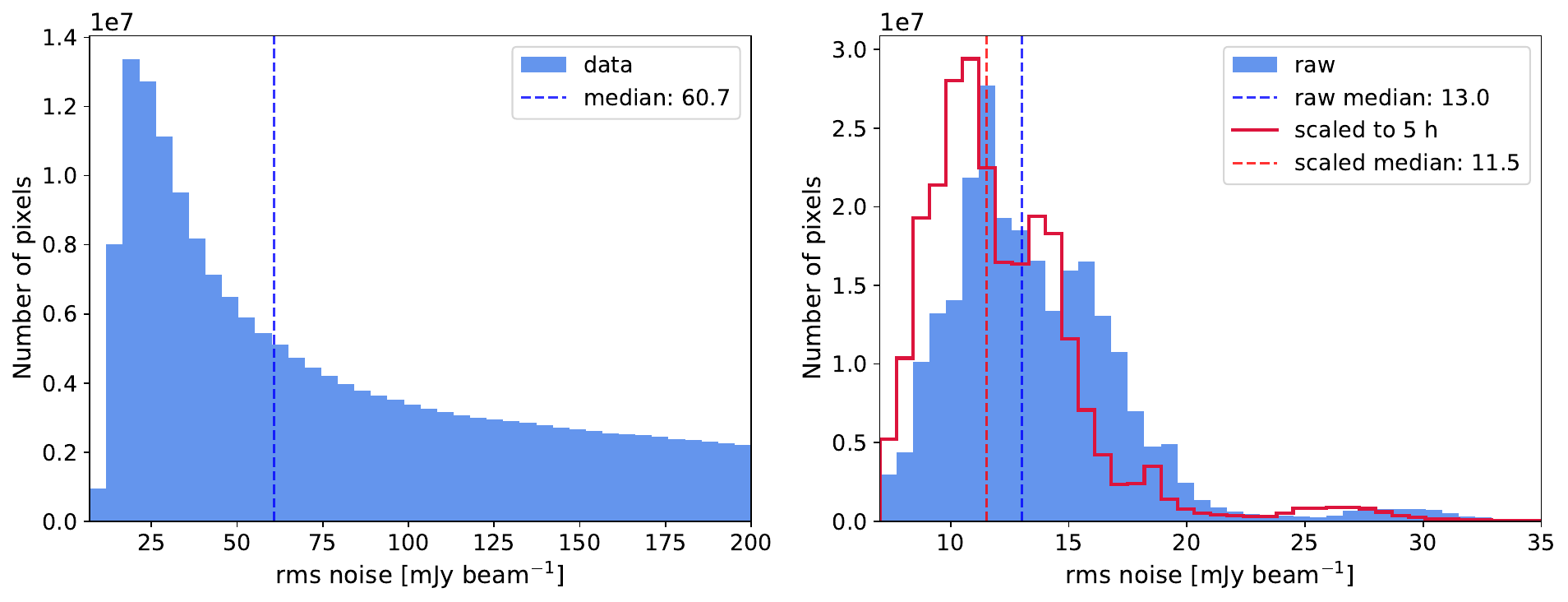}
    \caption{Rms noise distribution of pixels in the released images for the primary-beam-corrected products (left) and the non-primary-beam-corrected products (right). For the non-primary-beam-corrected images, both the raw distribution (blue) and the distribution scaled to a common effective integration time of 5 hours (red) are shown. The median noise is 60.7 mJy beam$^{-1}$ for the primary-beam-corrected images, while for the non-primary-beam-corrected images, the median decreases from 13.0 mJy beam$^{-1}$ in the raw distribution to 11.5 mJy beam$^{-1}$ after rescaling.}
    \label{fig:rms_hist}
\end{figure*}

We examined the rms noise distribution in primary-beam-corrected and uncorrected images by constructing histograms of values in the \texttt{PyBDSF} rms maps (\autoref{fig:rms_hist}). The primary-beam-corrected histogram (\autoref{fig:rms_hist} left) has a long tail towards high noise values, caused by the increase of noise at distances larger than 8$^\circ$ from the pointing centre.

To isolate calibration effects, we consider the non-primary-beam-corrected rms distribution (\autoref{fig:rms_hist} right, blue). This distribution shows a very broad peak, driven by variations in integration time and the fraction of data flagged during preprocessing. For example, a 5-hour observation with 30\% of the data flagged has a noise level comparable to that of an idealised 3.5-hour observation in which no data are flagged. To account for this, we rescale the rms values to a common effective integration time. For each field, the effective observing time is defined as
\begin{equation} t_{\mathrm{eff}} = t_{\mathrm{obs}}(1-f_{\mathrm{flag}}),
\end{equation}
where $t_{\mathrm{obs}}$ is the total integration time and $f_{\mathrm{flag}}$ is the fraction of flagged data. Assuming thermal noise behaviour, the measured rms is rescaled to a reference depth of 5 hours according to
\begin{equation}
\sigma_{5\,\mathrm{h}} = \sigma_{\mathrm{meas}} \sqrt{\frac{t_{\mathrm{eff}}}{5\,\mathrm{h}}},
\end{equation}
where $\sigma_{\mathrm{meas}}$ is the rms measured in the original image. The resulting corrected histogram (\autoref{fig:rms_hist} right, red) shows the main distribution narrowing, indicating that differences in effective exposure mainly cause the broadening. The remaining high-noise tail is associated with regions around bright sources. A secondary excess around $\sim$19~mJy is linked to a small number of fields, such as P68+23, P300+21, and P246+74, containing extremely bright central sources and affected by poor ionospheric conditions. After rescaling to a common effective integration time, the median central rms is 11.5~mJy beam$^{-1}$, close to the expected thermal noise of 10~mJy beam$^{-1}$.

\begin{figure}[!hbt]
    \centering
    \includegraphics[width=0.9\linewidth]{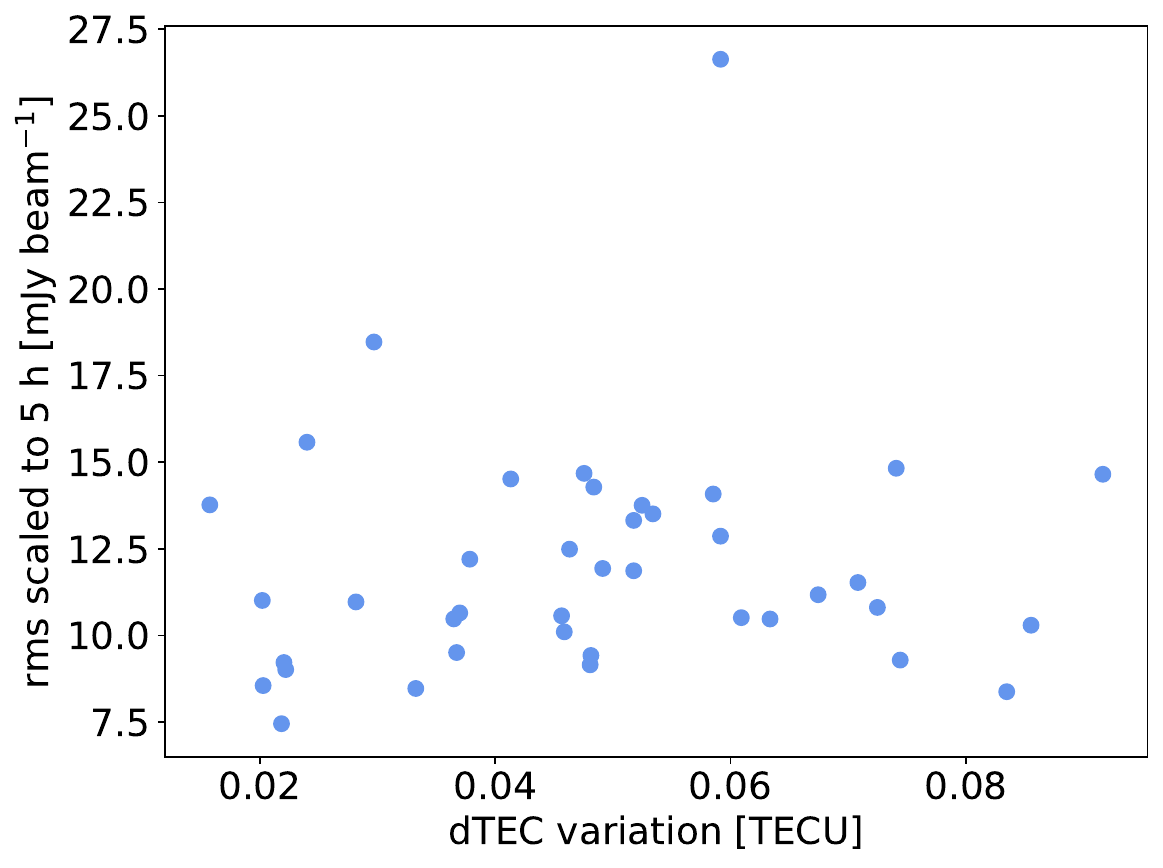}
    \caption{Central rms noise, scaled to an effective integration time of 5 hours, as a function of the dTEC variability for each field. The highest-rms outliers are associated with extremely bright sources and their residual calibration artefacts.}
    \label{fig:rms_dtec}
\end{figure}

To investigate whether the remaining field-to-field variation in image noise is related to ionospheric conditions, we compared the 5-hour-scaled rms with the variability of the differential total electron content (dTEC) derived from the calibrator solutions. For each one-hour observing block, we calculated the standard deviation of the dTEC solutions and averaged it over the five most distant Dutch stations. For fields composed of multiple observing blocks, the resulting values were combined using an integration-time-weighted average. As shown in \autoref{fig:rms_dtec}, no clear overall correlation is observed between the scaled rms and the dTEC variability. The highest-rms outliers correspond to fields containing extremely bright sources, whose residual calibration artefacts increase the noise across the image.

\subsection{Astrometric precision and accuracy}\label{sec:results_2}

We evaluate the astrometric precision and accuracy of our catalogue by cross-matching it with LoTSS DR3 \citep{shimwell2026lofar}. We use the signal-to-noise-dependent criterion from Sect.~\ref{sec:results_0} to select compact sources in our catalogue, and match them to LoTSS using the same one-to-one nearest-neighbour procedure as described there. For this comparison, LoTSS is used only as the astrometric reference catalogue, so we do not impose the additional \texttt{S\_Code = 'S'} selection on the LoTSS sources.

\begin{figure}[!htp]
    \centering
    \includegraphics[width=0.9\linewidth]{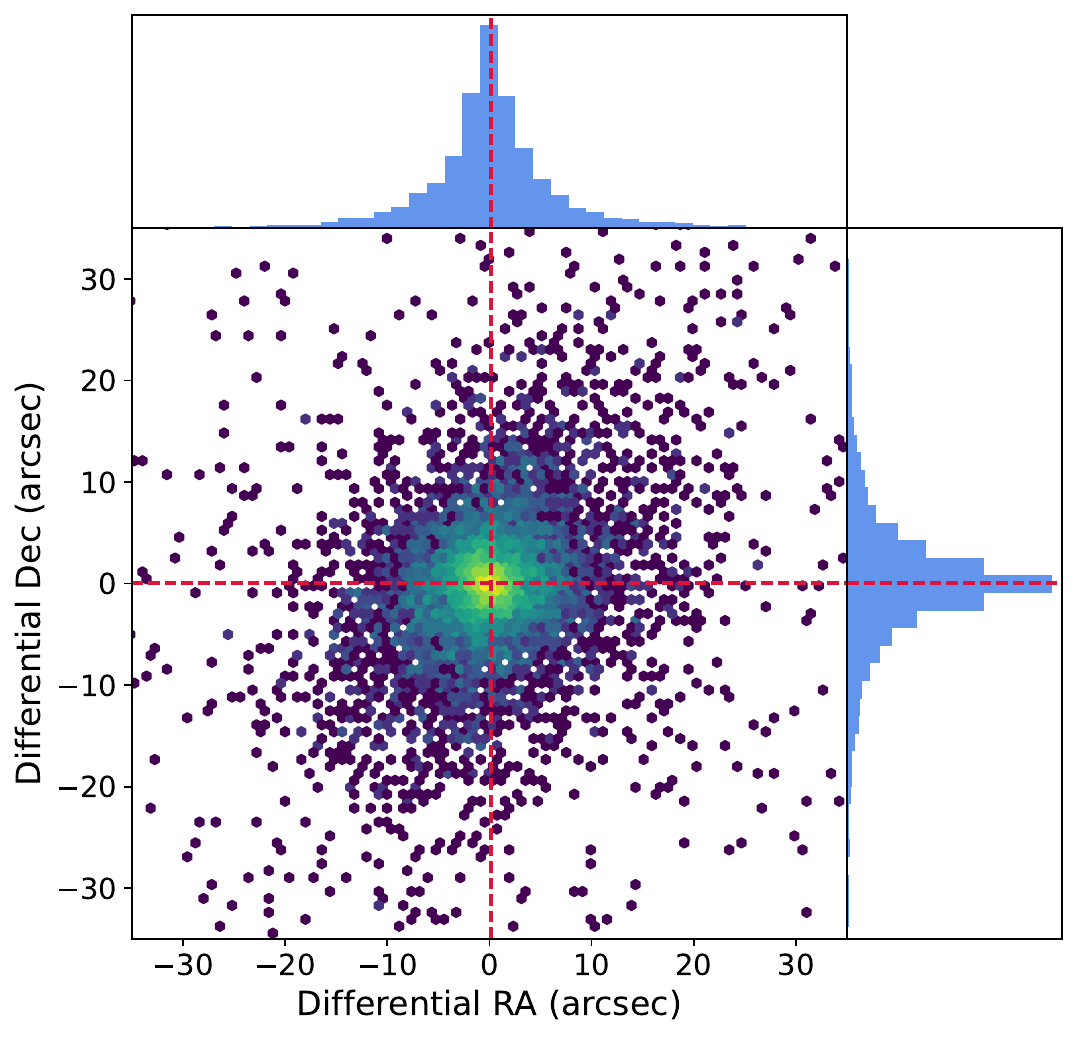}
    \caption{Residual astrometric offsets in right ascension and declination for the matched source sample. The dashed lines indicate the mean offsets, while the marginal histograms show the one-dimensional distributions. The astrometric precision is $\sigma_{\mathrm{RA}} = 8.70\arcsec$ and $\sigma_{\mathrm{Dec}} = 9.20\arcsec$, and the mean offsets are $\langle \Delta \mathrm{RA} \rangle = 0.16\arcsec$ and $\langle \Delta \mathrm{Dec} \rangle = 0.06\arcsec$.}
    \label{fig:astrometry}
\end{figure}

The resulting offset distribution is shown in \autoref{fig:astrometry}. We find mean offsets of $\langle \Delta \mathrm{RA} \rangle = 0.16\arcsec$ and $\langle \Delta \mathrm{Dec} \rangle = 0.06\arcsec$, both lower than 0.4\% of the synthesised beam, indicating that the per-field corrections successfully remove systematic astrometric shifts. The residual scatter provides an estimate of the astrometric precision, giving $\sigma_{\mathrm{RA}} = 8.70\arcsec$ and $\sigma_{\mathrm{Dec}} = 9.20\arcsec$.

\subsection{Flux density scale}\label{sec:results_3}

To study the flux-density scale, we adopt a nearest-neighbour cross-matching approach similar to that described in Sect.~\ref{sec:results_2}. We select only compact, isolated LoDeSS sources with $S/N > 10$. These sources are first matched to LoTSS using the procedure described above, after which the matched sample is cross-matched with the NVSS and VLSSr catalogues. We use these catalogues because they provide broad sky coverage over the LoDeSS fields, contain large numbers of sources, and span a wide frequency range, from 1.4 GHz for NVSS down to 73.8 MHz for VLSSr. LoLSS was not used in this analysis because the current LoLSS public release overlaps only marginally with the non-contiguous LoDeSS DR1 fields. A direct comparison between LoDeSS and LoLSS will be revisited in the final LoDeSS data release, once the full LoLSS release is available.

\begin{figure}
    \centering
    \includegraphics[width=0.95\linewidth]{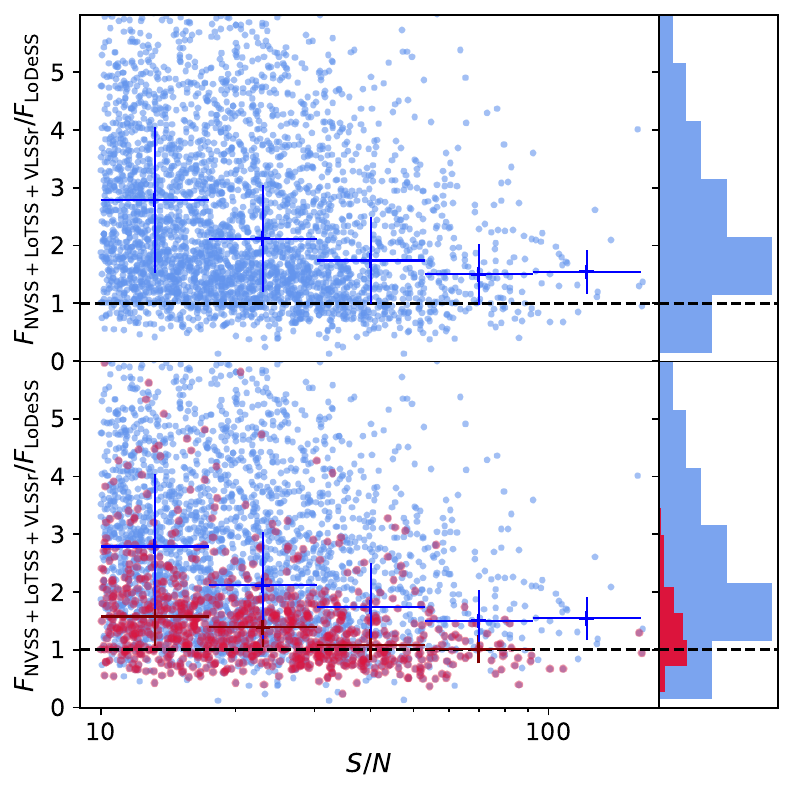}
    \caption{Ratio between the flux densities predicted at 23~MHz and those measured by LoDeSS. The predicted values are obtained by fitting a power law to the NVSS, LoTSS, and VLSSr measurements and evaluating the fit at 23~MHz. The top panel shows all matched sources, while the bottom panel highlights the high-quality fields in red. The black dotted line indicates a ratio of unity, corresponding to agreement between the predicted and measured flux densities. The crosses show binned median values, with blue corresponding to the full sample and red to the high-quality subset.}
    \label{fig:flux}
\end{figure}

Following the cross-matching, we fit a single power law to the NVSS, LoTSS, and VLSSr flux-density measurements, equivalent to a linear fit in log--log space. We then extrapolate this fit to the LoDeSS central frequency of 23 MHz and compare the predicted flux densities with the measured LoDeSS values. The resulting distribution is shown in the top panel of \autoref{fig:flux}.

The full sample shows both an offset from unity and substantial scatter. The median ratio between the predicted and measured flux densities is $R_{flux}=2.23$, with a semi-interquartile range ${\rm SIQR} = 1.15$. We first checked whether an overall calibration-scale error could cause this offset. The adopted calibrator models are consistent with \citet{scaife2012broad}, and the calibrated flux densities of the calibrator sources agree with the expected model values. This indicates that the residual offset is driven by in-field effects rather than by an incorrect global flux-density scale.

We attribute both the offset and the scatter primarily to residual ionospheric effects, which can cause decorrelation. The fields in this release were selected around exoplanet targets rather than based on observing conditions and therefore span a wide range of ionospheric quality. To illustrate this, we highlight in the bottom panel of \autoref{fig:flux} a subset of fields with comparatively stable ionospheric behaviour, selected by visual inspection of the phase solutions. For these fields, the scatter is reduced and the median flux-density ratio shifts closer to unity, from $R_{flux}=2.23$ for the full sample to $R_{flux}=1.39$ for the high-quality subset. The corresponding scatter decreases to ${\rm SIQR} = 0.40$. This supports the interpretation that the dominant flux-density uncertainty is field-dependent, most likely due to decorrelation caused by residual ionospheric effects. We discuss these effects and possible strategies to improve the remaining fields in Sect.~\ref{sec:discussion}. Residual uncertainties in the primary-beam model may also contribute, particularly away from the pointing centre, but they are unlikely to be the dominant source of the observed field-to-field scatter.

As an additional check, we repeat the comparison using NVSS, LoTSS, and 8C (see plot in Appendix~\ref{sec:ap4}). Since 8C lies closer to the LoDeSS observing frequency, this comparison is less sensitive to extrapolation from high frequencies. However, because low-frequency spectral curvature and turnovers are expected for some sources, we use a quadratic polynomial in log--log space rather than a single power law. The 8C comparison contains fewer matched sources because of the lower source density and coarser angular resolution of the 8C catalogue, but it shows the same qualitative behaviour as the VLSSr-based comparison. For the full sample, we find a median ratio of $R_{flux}=2.33$ with ${\rm SIQR} = 1.25$. For the high-quality fields, the median ratio decreases to $R_{flux}=1.30$, and the scatter is reduced to ${\rm SIQR} = 0.39$. We therefore treat the 8C comparison as a consistency check rather than as the primary flux-scale diagnostic.

A small number of sources with elevated flux-density ratios remain even in the high-quality fields. These may correspond to genuine spectral turnovers or peaked-spectrum sources, which become prominent at low frequencies \citep{callingham2017extragalactic}.

The catalogue flux-density uncertainties reported by PyBDSF include the local statistical uncertainty but not systematic effects from ionospheric smearing, residual beam-model errors, or the external flux-density scale. Based on the high-quality 8C comparison, we recommend adopting an additional systematic flux-density uncertainty of $39\%$ for bright, compact, isolated sources, corresponding to the SIQR of the flux-density ratio. This empirical value includes contributions from both LoDeSS systematics and the external comparison, but provides a practical uncertainty estimate for comparisons with other surveys. This additional 39\% systematic uncertainty is not included in the flux-density uncertainty columns of the released catalogues, which contain only the statistical uncertainties reported by PyBDSF.

\begin{figure}
    \centering
    \includegraphics[width=\linewidth]{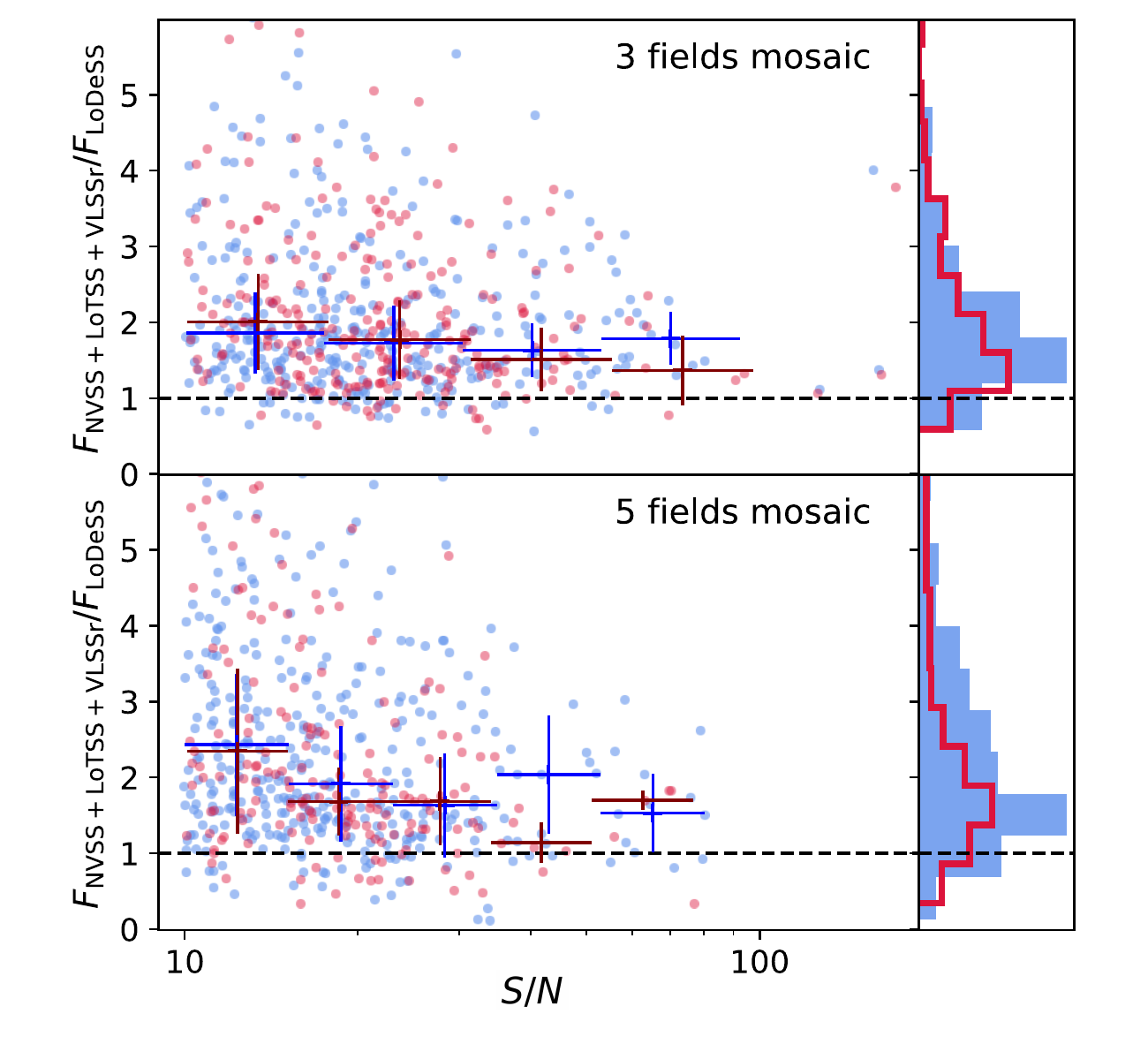}
    \caption{Flux-density ratios before (blue) and after (red) mosaicing for the mosaics obtained from three fields (top) and five fields (bottom).}
    \label{fig:mosaic_flux}
\end{figure}

As a further test of whether mosaicing helps with the flux density scatter, we repeated the NVSS--LoTSS--VLSSr comparison for the two regions covered by mosaics, one combining three fields and the other five. For each region, we compared the flux densities predicted at 23~MHz with the measurements obtained from the individual pointings and from the corresponding mosaic. The results are shown in Fig.~\ref{fig:mosaic_flux}.

For the three-field region, the median flux-density ratio changes from $R_{\rm flux}=1.79$ for the individual pointings to $R_{\rm flux}=1.78$ for the mosaic, while the scatter decreases slightly from ${\rm SIQR}=0.53$ to ${\rm SIQR}=0.52$. The modest improvement is consistent with two of the three contributing fields already having comparatively good ionospheric quality.

For the five-field region, the improvement is more pronounced. The median ratio decreases from $R_{\rm flux}=2.06$ to $R_{\rm flux}=1.83$, while the scatter decreases from ${\rm SIQR}=0.95$ to ${\rm SIQR}=0.70$. This larger change is consistent with the stronger ionospheric degradation present in several of the contributing fields, whose influence is reduced in the weighted mosaic. These results show that mosaicing can improve the internal consistency of the flux-density scale, particularly when the contributing pointings have heterogeneous image quality.

\subsection{Stokes V leakage}\label{sec:results_4}

To estimate the instrumental leakage into Stokes~V, we followed a method similar to that of \citet{callingham2023v} and \citet{shimwell2022lofar}. A representative example of the Stokes~I and Stokes~V images used for this assessment is shown in Appendix~\ref{sec:ap3}, where the residual leakage can be visually identified around bright Stokes~I sources. We selected sources from our catalogue that are isolated within $90\arcsec$, classified as point-like (Sect.~\ref{sec:results_0}), and detected with $S/N > 10$. We further restricted the sample to sources lying within the FWHM of the primary beam. For each selected source, we measured the Stokes~V flux density in the corresponding primary-beam-corrected Stokes~V image. Specifically, we searched for the maximum absolute Stokes~V emission ($S_V$) within a radius of $45\arcsec$ from the Stokes~I source position, which matches the angular resolution of our data. The total Stokes~I flux density, $S_I$, and the local Stokes~I rms noise, $\sigma_I$, were taken from the PyBDSF source catalogues.

\begin{figure}
    \centering
    \includegraphics[width=0.9\linewidth]{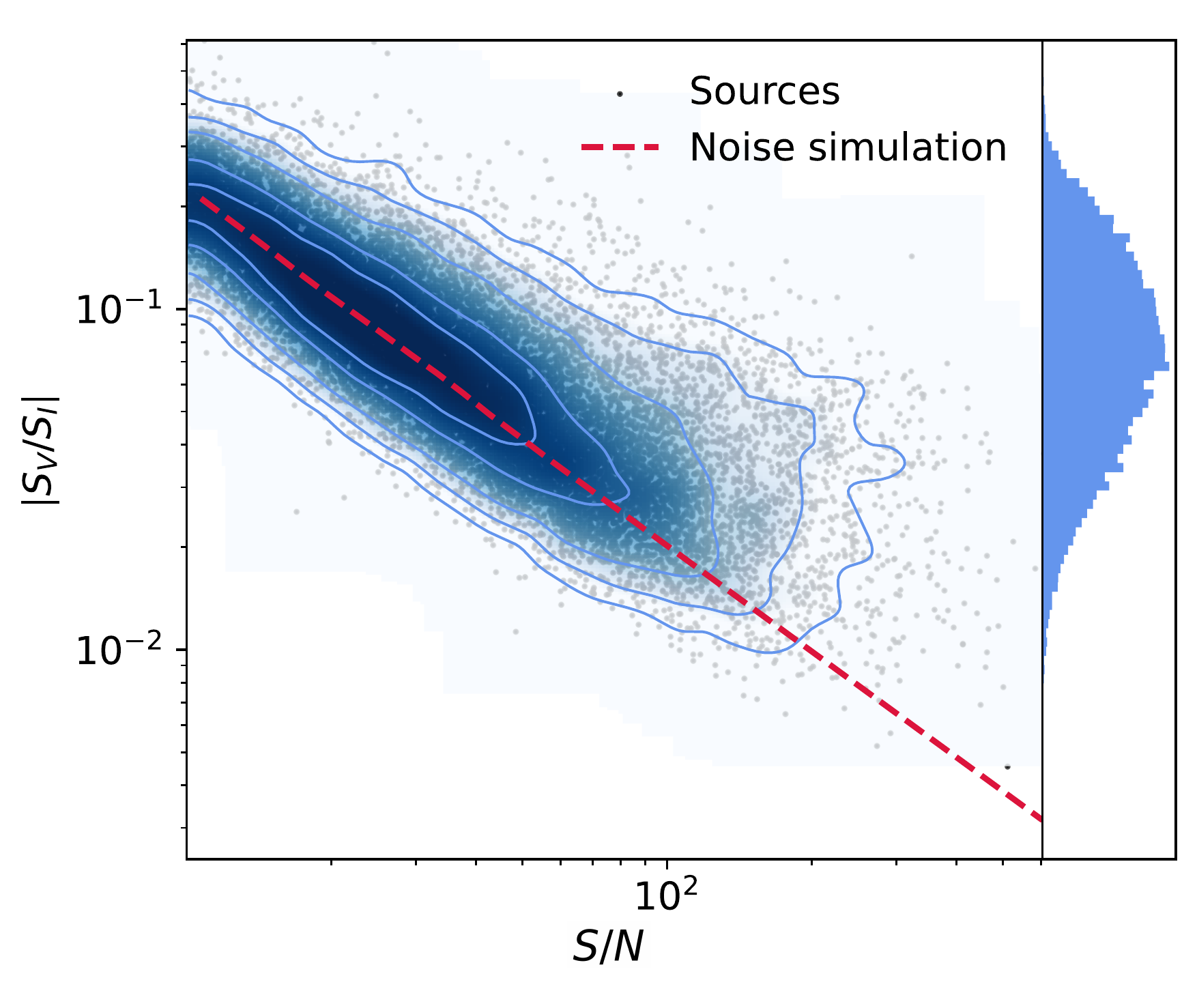}
    \caption{Stokes V~leakage as a function of the S/N for isolated, point-like sources. The red line shows the expectation from Gaussian noise.}
    \label{fig:leakage}
\end{figure}

To quantify the leakage expected from noise alone, we performed noise-only simulations. For each source, we estimated the local Stokes~V image rms and generated Gaussian random noise within an aperture of the same size as that used for the Stokes~V measurement. We then measured the maximum absolute value in each case, providing the expected behaviour of $|S_V|/S_I$ in the absence of leakage. The results are shown in \autoref{fig:leakage}, where the red dashed line indicates the Gaussian noise expectation.

The observed measurements lie systematically above the noise-only relation. Since the sample is restricted to bright, isolated, point-like Stokes~I sources, we interpret this excess as evidence for instrumental Stokes~$I \rightarrow V$ leakage. We, therefore, model the trend as
\begin{equation}
    \log \left| \frac{S_V}{S_I} \right| = A \,\log \left( \frac{S_I}{\sigma_I} \right) + B.
\end{equation}
From this fit, we derive a characteristic leakage fraction of $\mathrm{1.71 \pm 0.09\%}$. This is comparable to the Stokes~V leakage reported for LoLSS DR2, which also uses the LOFAR LBA dipoles, where typical values are $\sim1$--$2\%$ before calibration, and the median post-calibration leakage scatter is $\sim0.25\%$ (Edler et al., in prep.). This is consistent with the difficulty of polarisation calibration at decametre wavelengths. We therefore consider circularly polarised sources with fractional circular polarisation above 10\% to be robust against leakage contamination.

\section{Discussion}\label{sec:discussion}

\begin{figure}
    \centering
    \includegraphics[width=1.0\linewidth]{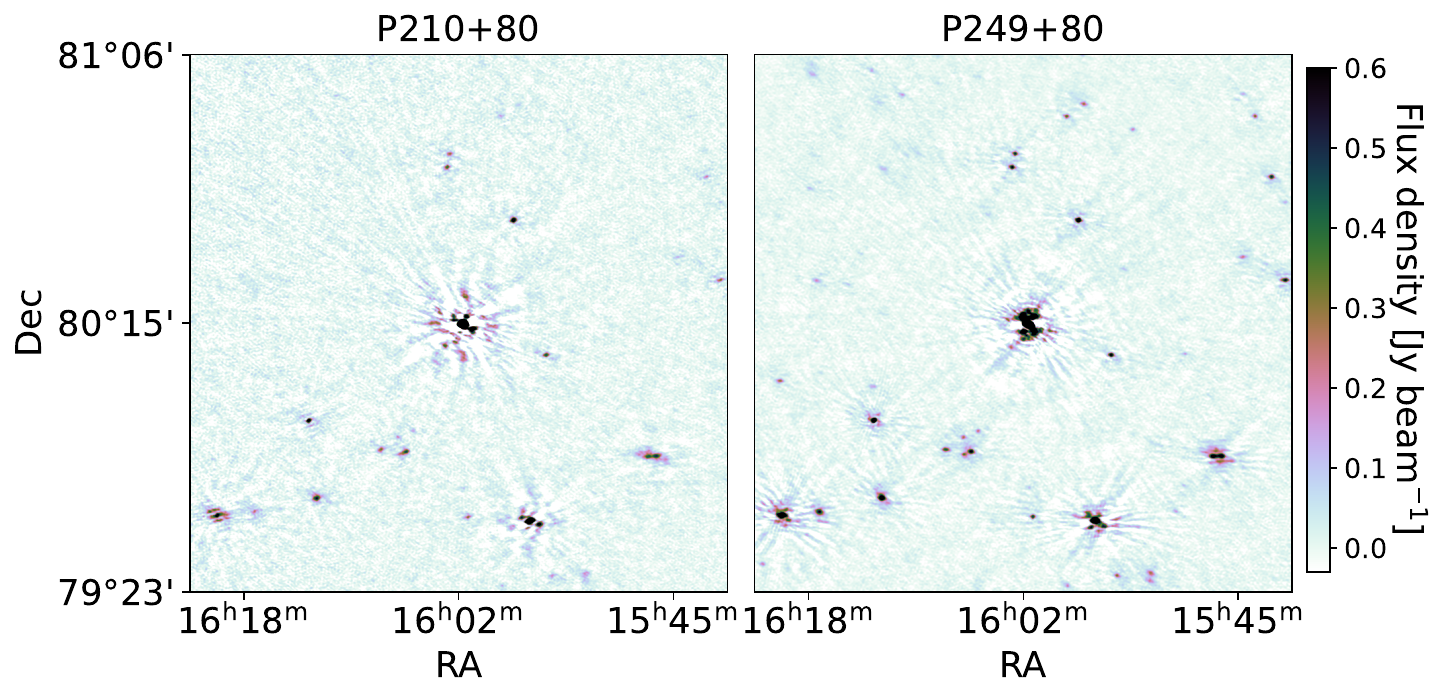}
    \caption{Comparison of the same sources in a field with moderate (left) and good (right) ionospheric conditions. The source at the phase centre coincides with the centre of the calibration facet in both observations, and has flux densities of 21.7 Jy beam$^{-1}$ (left) and 22.8 Jy beam$^{-1}$ (right), corresponding to a difference of $\sim$5\%, as expected \citep{de2023lofar}. At increasing distances from the facet centre, sources in the left panel appear progressively fainter compared to their counterparts in the right panel, indicating stronger ionospheric smearing. Some of the faintest sources disappear entirely in the left panel.}
    \label{fig:fields_comp}
\end{figure}

Observations at decametre wavelengths are strongly affected by the ionosphere, which is the dominant limitation on image quality and flux-density-scale accuracy in this data release. This effect is particularly pronounced for sources far from the facet centre, where residual phase errors lead to significant smearing and decorrelation. Since calibration facets are centred on the brighter sources, fainter sources are more likely to lie near facet edges and are therefore more strongly affected. This is reflected in the flux-scale analysis (Sect.~\ref{sec:results_3}), where restricting the analysis to high-quality fields significantly reduces the scatter. We confirmed this by comparing two adjacent fields observed under different ionospheric conditions.   Sources located near the centres of calibration facets show consistent, and therefore more reliable, flux densities between fields, while those further away exhibit increased scatter, with some sources disappearing altogether in the field affected by poorer ionospheric conditions (\autoref{fig:fields_comp}). The two fields also show a consistent trend in diffractive scale and spatial decorrelation angle: fields affected by poorer ionospheric conditions have smaller diffractive scales, and phase errors become significant at smaller angular separations from the facet centre \citep{10.1117/12.3101297}.

\begin{figure*}
    \centering
    \includegraphics[width=0.8\linewidth]{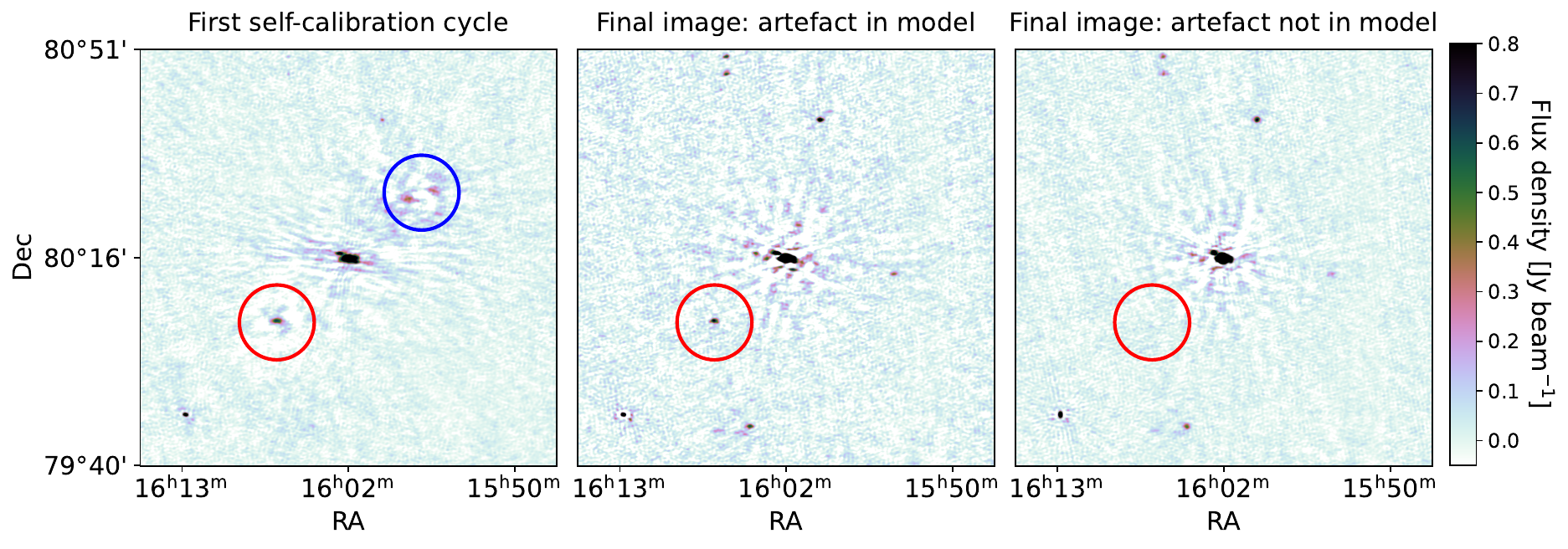}
    \caption{First self-calibration cycle, where the artefact appears (left), the final image with the artefact included in the calibration model (middle), and the final image after removal of the artefact from the model (right). The artefact is marked with a red circle. The feature is identified as an artefact based on its negative counterpart, visible opposite the bright source in the left panel (blue circle). This structure is located at a distance of $\sim 7^\circ$ from the phase centre.}
    \label{fig:artefact}
\end{figure*}

Low S/N can also introduce atypical calibration artefacts. In some cases, model components introduced during self-calibration do not converge to the true sky model or are not fully removed, leading to persistent imaging artefacts, as illustrated in \autoref{fig:artefact}. Although the affected facets remain stable during subsequent self-calibration cycles, the spurious model components can persist even after additional cycles. Future improvements will therefore require more physically constrained calibration schemes, for example, by exploiting the frequency dependence of ionospheric phases or by imposing spatial correlations between solutions in neighbouring directions \citep{albert2020probabilistic}.

In the example shown in \autoref{fig:artefact}, the spurious source was identified because it appeared as an extremely steep-spectrum source with no counterpart in LoTSS or LoLSS. A subsequent search of the available multi-wavelength surveys and catalogues in VizieR, including radio, optical, infrared, and X-ray data, likewise revealed no counterpart at its position. The source was also absent from overlapping LoDeSS fields. The artefact was therefore manually removed from the self-calibration model, after which it disappeared from the resulting image. All fields in the present data release were then visually inspected for similarly steep-spectrum sources without multi-wavelength counterparts, but no additional cases were found. The single case occurred in a field with moderate ionospheric conditions, and it is likely related to a combination of low S/N and limitations of the input sky model. Nevertheless, this illustrates an additional challenge that must be considered when processing large datasets at these frequencies.

Another important source of uncertainty is the primary-beam model. At present, the beam response of the LOFAR LBA at decametre wavelengths is not well constrained, and inaccuracies in the beam model likely contribute to the elevated Stokes~V leakage reported in Sect.~\ref{sec:results_4}. Improved LBA beam models have been developed and are expected to reduce these uncertainties, but they still require further validation and an official release before they can be routinely used in calibration.

The observed flux-density scatter (Sect.~\ref{sec:results_3}) can be mitigated through mosaicing. The survey design includes substantial overlap between adjacent pointings, allowing multiple independent measurements of the same regions of sky. Our tests on the two available mosaics show that the scatter is reduced when fields are combined, with the largest improvement obtained for the region containing several pointings affected by poorer ionospheric conditions. This indicates that mosaicing is most effective when the weighting accounts not only for the distance from the pointing centre, but also for field-dependent image quality. Users should nevertheless remain aware that substantial field-dependent flux-density scatter is still present outside the tested mosaic regions.

One way to reduce the flux-density scatter in the upcoming low-frequency surveys is through the ongoing LOFAR~2.0 upgrade. In the upgraded system, the LBA will be able to use all 96 antennas per station, compared to the 48 antennas used in the current configuration. At $\sim$50~MHz, this corresponds to a factor of two increase in sensitivity, while at decametre wavelengths the improvement is expected to be somewhat smaller due to overlapping effective areas of the innermost dipoles. Nevertheless, the increased sensitivity will reduce the image noise and increase the number of in-field sources that are sufficiently bright to constrain direction-dependent calibration. This should make it possible to define more calibration facets, which in turn is expected to reduce residual ionospheric smearing and improve the flux-density consistency across the field.

\section{Conclusion} \label{sec:conclusion}

In this work, we have presented the first data release of the LOFAR Decametre Sky Survey (LoDeSS), including calibrated images and source catalogues in the $15-30$~MHz regime. We have shown that, despite the severe challenges posed by the ionosphere, radio-frequency interference, and calibration limitations, it is possible to obtain sensitive, wide-field images at these frequencies, reaching typical noise levels of $\sim$ 12 $\rm mJy \ beam^{-1}$. This release improves on previous wide-area decametre surveys, such as the 8C survey at 38\,MHz, by approximately an order of magnitude in both depth and angular resolution, and significantly expands on the two-pointing demonstration presented by \citet{groeneveld2024characterization}.

The data quality does not reach that of higher-frequency LOFAR surveys, which is expected given the extreme observing regime. LoDeSS pushes radio observations to significantly lower frequencies, opening a new part of the electromagnetic spectrum. Working in this regime is inherently difficult and comes with additional challenges, but it is also uniquely informative. In this context, these observations are not only complementary to higher-frequency surveys but will remain relevant even in the era of future facilities, as this frequency range is not accessible to instruments such as the Square Kilometre Array \citep[SKA;][]{braun2019anticipated}.

This first data release covers 15\% of the full survey. A final data release is planned, in which mosaicing of all fields will be applied. This will provide improved sensitivity, more uniform noise properties, and a more reliable flux density scale across the full survey area. These improvements will further enhance the scientific value of LoDeSS.

\section*{Data availability}

The LoDeSS DR1 source and Gaussian-component catalogues, primary-beam-corrected Stokes~I and Stokes~V images, and the mosaiced Stokes~I images presented in this article are available in electronic form at the CDS via anonymous ftp to \texttt{cdsarc.u-strasbg.fr} or via \url{http://cdsweb.u-strasbg.fr/cgi-bin/qcat?J/A+A/}. The data products are also available through the LOFAR Surveys website at \url{https://lofar-surveys.org/lodess.html}.

\begin{acknowledgements}
      CMC and HKV acknowledge funding from the European Research Council via the starting grant `STORMCHASER' (grant number 101042416). CG and FdG acknowledge support from the ERC Consolidator Grant ULU 101086378. JRC acknowledges funding from the European Union via the ERC grant Epaphus (project number: 101166008). This paper is based on data obtained with the LOFAR telescope (LOFAR-ERIC) under project code LT16\_014. LOFAR \citep{VanHaarlem2013} is the Low Frequency Array designed and constructed by ASTRON. It has observing, data processing, and data storage facilities in several countries, that are owned by various parties, each with their own funding sources, and that are collectively operated by the LOFAR European Research Infrastructure Consortium (LOFAR-ERIC) under a joint scientific policy. This work made use of the Python packages \texttt{matplotlib} \citep{mpl} to generate the figures and \texttt{numpy} \citep{np} for computations. CMC acknowledges the use of ChatGPT for grammar correction and phrasing improvements.
\end{acknowledgements}

\bibliographystyle{aa} 
\bibliography{references}

\FloatBarrier 
\clearpage
\begin{appendix}

\onecolumn

\section{List of LoDeSS processed fields}\label{sec:ap1}

\begin{table}[!ht]
\caption{Additional information about all processed fields.}
\label{tab:fields}
\centering
\begin{tabular}{lcccccc}
\hline\hline
Field & RA [deg] & Dec [deg] & Obs. duration [h] & Flagged [\%] & RMS [mJy~beam$^{-1}$] & Quality \\
\hline
P8+40   & 8.785   & 40.861 & 4   & 9.5  & 17.22 & mediocre \\
P17+40  & 17.026  & 40.999 & 3   & 23.0 & 12.32 & poor \\
P21+29  & 21.448  & 29.012 & 3   & 33.0 & 16.24 & mediocre \\
P36+35  & 36.525  & 35.297 & 5   & 9.6  & 11.06 & good \\
P57+47  & 57.856  & 47.711 & 5   & 7.8  & 8.82  & good \\
P61+23  & 61.963  & 23.658 & 5.5 & 16.1 & 14.87 & mediocre \\
P68+23  & 68.759  & 23.772 & 4   & 16.0 & 18.08 & mediocre \\
P78+29  & 78.640  & 29.971 & 6   & 18.0 & 13.88 & mediocre \\
P90+72  & 90.346  & 72.304 & 6   & 23.5 & 9.69  & mediocre \\
P100+30 & 100.231 & 30.332 & 5   & 13.4 & 9.69  & poor \\
P109+24 & 109.660 & 24.457 & 4   & 13.5 & 11.08 & mediocre \\
P114+30 & 114.668 & 30.574 & 5   & 15.6 & 11.40 & poor \\
P128+36 & 128.933 & 36.844 & 5.5 & 16.2 & 7.76  & good \\
P130+60 & 130.929 & 60.978 & 4   & 15.7 & 11.14 & good \\
P136+30 & 136.393 & 30.938 & 5   & 18.3 & 9.46  & mediocre \\
P136+36 & 136.717 & 36.974 & 3   & 37.3 & 15.36 & good \\
P139+79 & 139.091 & 79.052 & 5   & 23.4 & 12.58 & good \\
P143+25 & 143.915 & 25.031 & 5   & 18.0 & 13.18 & mediocre \\
P151+49 & 151.859 & 49.285 & 4   & 13.0 & 11.39 & good \\
P171+49 & 171.021 & 49.606 & 4   & 15.5 & 12.95 & good \\
P181+55 & 181.436 & 55.804 & 4.5 & 16.4 & 15.57 & good \\
P208+68 & 208.688 & 68.284 & 6   & 13.3 & 10.27 & poor \\
P210+80 & 210.172 & 80.211 & 5   & 10.8 & 11.44 & mediocre \\
P219+50 & 219.491 & 50.417 & 5   & 9.6  & 10.63 & good \\
P220+20 & 220.669 & 20.284 & 5   & 8.4  & 19.30 & mediocre \\
P222+74 & 222.379 & 74.494 & 6   & 20.8 & 11.46 & mediocre \\
P238+32 & 238.906 & 32.655 & 5   & 13.6 & 11.80 & good \\
P240+26 & 240.708 & 26.653 & 5   & 17.6 & 14.17 & mediocre \\
P246+74 & 246.234 & 74.889 & 5   & 10.2 & 20.27 & good \\
P249+80 & 249.259 & 80.845 & 5   & 11.0 & 12.93 & good \\
P264+86 & 264.991 & 86.543 & 8   & 14.5 & 8.79  & mediocre \\
P268+27 & 268.614 & 27.121 & 5   & 28.2 & 15.72 & mediocre \\
P271+39 & 271.200 & 39.227 & 5   & 28.4 & 14.02 & mediocre \\
P278+63 & 278.206 & 63.438 & 7   & 27.0 & 11.40 & mediocre \\
P288+51 & 288.787 & 51.577 & 5   & 35.2 & 13.13 & mediocre \\
P290+33 & 290.905 & 33.526 & 5   & 19.8 & 16.39 & poor \\
P300+21 & 300.615 & 21.624 & 5   & 17.0 & 29.23 & poor \\
P308+45 & 308.683 & 45.883 & 5   & 19.0 & 13.88 & mediocre \\
P317+27 & 317.734 & 27.944 & 5   & 26.4 & 16.92 & poor \\
P318+58 & 318.642 & 58.098 & 5   & 17.8 & 15.17 & poor \\
\hline
\end{tabular}
\tablefoot{
The noise is estimated from the non-primary-beam-corrected images because this is more representative of the residual calibration errors. The field-quality classification is described in Sect.~4.3.
}
\end{table}

\twocolumn

\section{Catalogue columns}\label{sec:ap2}

\begin{table}[!ht]
\centering
\caption{Relevant columns in the PyBDSF source catalogue.}
\label{tab:catalogue_columns}
\begin{tabular}{lll}
\hline
Column & Unit & Description \\
\hline
Source\_id     & --      & Unique source identifier \\
Isl\_id        & --      & Island identifier \\
RA             & deg     & Right ascension of source centre \\
E\_RA          & deg     & Uncertainty on right ascension \\
DEC            & deg     & Declination of source centre \\
E\_DEC         & deg     & Uncertainty on declination \\
Peak\_flux     & Jy beam$^{-1}$ & Peak flux density \\
E\_Peak\_flux  & Jy beam$^{-1}$ & Uncertainty on peak flux density \\
Total\_flux    & Jy      & Total flux density \\
E\_Total\_flux & Jy      & Uncertainty on total flux density \\
RA\_max        & deg     & Right ascension of peak position \\
DEC\_max       & deg     & Declination of peak position \\
Maj            & arcsec  & Beam major axis size \\
Min            & arcsec  & Beam minor axis size \\
PA             & deg     & Position angle \\
Isl\_rms       & Jy beam$^{-1}$ & Local rms noise of the island \\
S\_Code        & --      & Source classification flag (point-like or extended) \\
\hline
\end{tabular}
\tablefoot{
The flux-density uncertainty columns contain only the statistical uncertainties reported by PyBDSF. The additional 39\% systematic uncertainty discussed in Sect.~5.4 is not included.
}
\end{table}

\FloatBarrier
\clearpage

\section{Stokes image examples}\label{sec:ap3}

\begin{strip}
    \centering
    \includegraphics[width=0.94\textwidth]{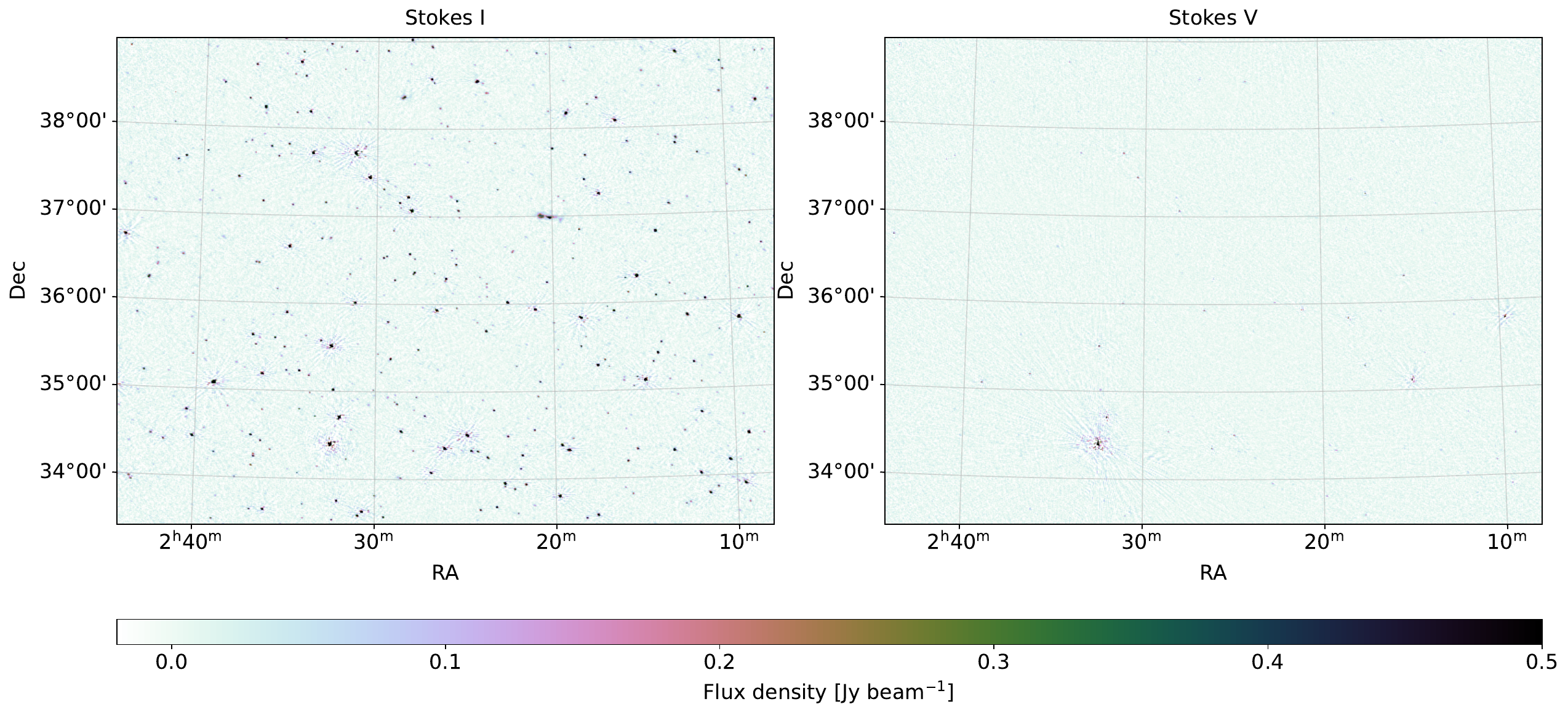}
    \captionof{figure}{Stokes~I (left) and Stokes~V (right) images of a zoomed-in region of the P36+35 field. The Stokes~V image shows residual leakage around the brightest Stokes~I sources.}
    \label{fig:stokes_iv_example}
\end{strip}

\section{Flux comparison with 8C catalogue}\label{sec:ap4}

\begin{figure}[!ht]
    \centering
    \includegraphics[width=0.95\linewidth]{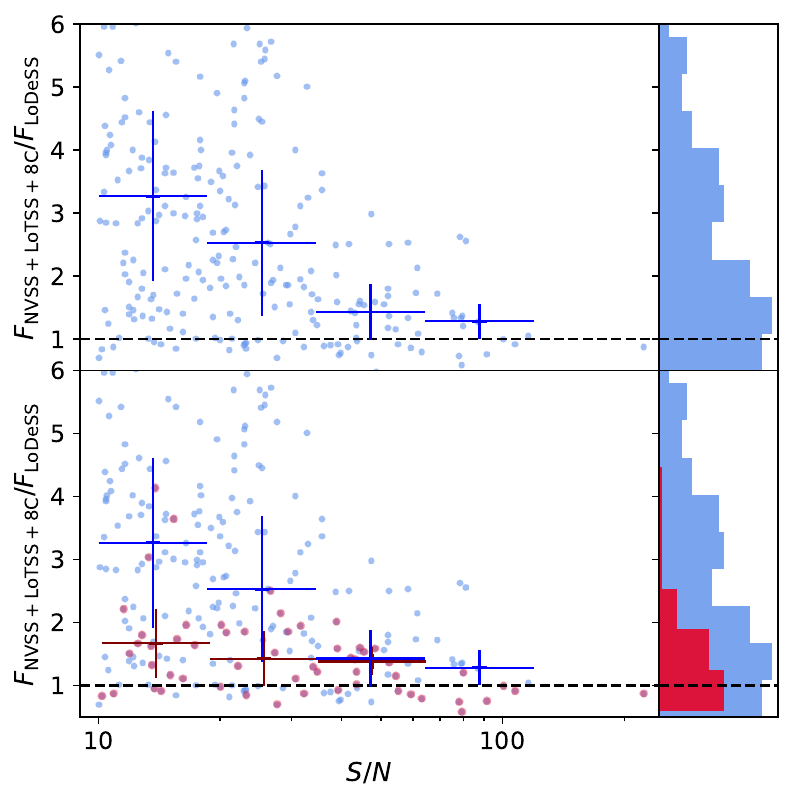}
    \caption{The ratio between the flux densities predicted at 23~MHz and those measured by LoDeSS. The predicted values are obtained by fitting a quadratic polynomial curve in log--log space to the NVSS, LoTSS, and 8C measurements and evaluating the fit at 23~MHz. The top panel shows all matched sources, while the bottom panel highlights the high-quality fields in red. The black dotted line indicates a ratio of unity, corresponding to agreement between the predicted and measured flux densities. The crosses show binned median values, with blue corresponding to the full sample and red to the high-quality subset.}
    \label{fig:8c_flux_ratio}
\end{figure}

\FloatBarrier

\end{appendix}
\end{document}